\documentclass[fleqn,usenatbib]{mnras}
\usepackage{newtxtext,newtxmath}
\usepackage[T1]{fontenc}
\usepackage{ae,aecompl}
\usepackage{graphicx}	
\usepackage{amsmath}	
\usepackage{amssymb}	
\usepackage{booktabs}
\usepackage{multirow}
\usepackage{ctable}
\usepackage{hyperref}
\title[Spectroscopic binaries in the Solar Twin Planet Search]{Spectroscopic binaries in the Solar Twin Planet Search program: from substellar-mass to M dwarf companions}

\author[L. A. dos Santos et al.]{Leonardo A. dos Santos,$^{1}$\thanks{E-mail: leonardoags@usp.br}
Jorge Mel\'endez,$^{1}$
Megan Bedell,$^{2}$
Jacob L. Bean,$^{2}$
\newauthor
Lorenzo Spina,$^{1}$
Alan Alves-Brito,$^{3}$
Stefan Dreizler,$^{4}$
Iv\'an Ram\'irez,$^{5}$
\newauthor
and Martin Asplund$^{6}$
\\
$^{1}$Universidade de S\~ao Paulo, Instituto de Astronomia, Geof\'isica e Ci\^encias Atmosf\'ericas, Rua do Mat\~ao 1226, S\~ao Paulo \\05508-090, Brazil\\
$^{2}$University of Chicago, Department of Astronomy and Astrophysics, 5640 S. Ellis Ave, Chicago, IL 60637, USA\\
$^{3}$Universidade Federal do Rio Grande do Sul, Instituto de F\'isica, Av. Bento Gon\c{c}alves 9500, Porto Alegre, RS, Brazil\\
$^{4}$University of G\"ottingen, Institut f\"ur Astrophysik, Germany\\
$^{5}$Tacoma Community College, Washington, USA\\
$^{6}$The Australian National University, Research School of Astronomy and Astrophysics, Cotter Road, Weston, ACT 2611, Australia\\
}

\date{Accepted XXX. Received YYY; in original form ZZZ}

\pubyear{2017}

\begin{document}
\label{firstpage}
\pagerange{\pageref{firstpage}--\pageref{lastpage}}
\maketitle

\begin{abstract}
Previous studies on the rotation of Sun-like stars revealed that the rotational rates of young stars converge towards a well-defined evolution that follows a power-law decay. It seems, however, that some binary stars do not obey this relation, often by displaying enhanced rotational rates and activity. In the Solar Twin Planet Search program, we observed several solar twin binaries, and found a multiplicity fraction of $42\% \pm 6\%$ in the whole sample; moreover, at least three of these binaries (HIP 19911, HIP 67620 and HIP 103983) clearly exhibit the aforementioned anomalies. We investigated the configuration of the binaries in the program, and discovered new companions for HIP 6407, HIP 54582, HIP 62039 and HIP 30037, of which the latter is orbited by a $0.06$ M$_\odot$ brown dwarf in a 1-month long orbit. We report the orbital parameters of the systems with well-sampled orbits and, in addition, the lower limits of parameters for the companions that only display a curvature in their radial velocities. For the linear trend binaries, we report an estimate of the masses of their companions when their observed separation is available, and a minimum mass otherwise. We conclude that solar twin binaries with low-mass stellar companions at moderate orbital periods do not display signs of a distinct rotational evolution when compared to single stars. We confirm that the three peculiar stars are double-lined binaries, and that their companions are polluting their spectra, which explains the observed anomalies.
\end{abstract}




\section{Introduction}

It is known that at least half of the stars in the Galaxy are multiple systems containing two or more stars orbiting each other \citep{2001ASPC..229...91K, 2017ApJ...836..139F}, thus in many surveys and large samples of stars, binaries are ubiquitous. This is in contrast with the Sun, which is a single star, and attempts to find a faint stellar companion orbiting it rendered no results thus far \citep[e.g.,][]{2014ApJ...781....4L}. Many studies avoid contamination by binaries in their samples, the main reasons being because we do not understand well how binaries evolve and how the presence of a companion affects the primary star. However, with the development of instruments with higher spatial and spectral resolution and coronagraphs, it is now possible to better probe the secondary component of such systems.

\defcitealias{2014A&A...572A..48R}{Paper~I}
\defcitealias{2015A&A...581A..34B}{Paper~II}
\defcitealias{2016A&A...590A..32T}{Paper~III} \defcitealias{2016A&A...592A.156D}{Paper~IV}
\defcitealias{2017A&A...597A..34M}{Paper~V}

We have been carrying out a radial velocity planet search focused on solar twins using HARPS \citep[][hereafter Papers I, II, III, IV and V, respectively]{2014A&A...572A..48R, 2015A&A...581A..34B, 2016A&A...590A..32T, 2016A&A...592A.156D, 2017A&A...597A..34M}. The definition of solar twin we use is a star with stellar parameters inside the ranges $5777 \pm 100$ K, $4.44 \pm 0.10$ dex(cgs) and $0.0 \pm 0.1$ dex, respectively, for $T_{\mathrm{eff}}$, $\log{g}$ and [Fe/H]. In total, 81 solar twins\footnote{\footnotesize{Some of the stars in our sample do not fit the strict definition of solar twins because one or more parameters are off the definition intervals, but they are still very close solar analogues.}} were observed on HARPS. As part of our survey we previously identified 16 clear spectroscopic binaries (SB) \citepalias{2016A&A...592A.156D}. We report here the identification of four additional SBs (HIP 14501, HIP 18844, HIP 65708 and HIP 83276) and the withdrawal of HIP 43297 and HIP 64673, which are unlikely to host stellar-mass companions, bringing the number of solar twin SBs to 18. Most of these SBs are single-lined -- they do not contain a second component in their spectral lines --, meaning that their companions are either faint stars or located outside the $\sim$$1\arcsec$ aperture of the HARPS spectrograph. We confirm that there are three solar twins with spectra contaminated by a relatively bright companion (see discussion in Section \ref{peculiar}). In our sample there are an additional 18 visual binaries\footnote{\footnotesize{We define as visual companions those with separations larger than $1\arcsec$}} or multiple systems, of which HIP 6407 and HIP 18844 have wide companions \citep[see table 5 in][]{2014AJ....147...86T} as well as the spectroscopic companions reported here.

In \citetalias{2016A&A...592A.156D} we saw that the single or visual binary solar twins display a rotational evolution that can described with a relation between stellar age $t$ and rotational velocity $v_{\mathrm{rot}}$ in the form of a power law plus a constant: $v_{\mathrm{rot}} = v_{\mathrm{f}} + m\ t^{-b}$, where $v_{\mathrm{f}}$, $m$ and $b$ are free parameters to be fit with observations. This relation is explained by loss of angular momentum due to magnetized winds \citep[e.g.,][]{1984RSPTA.313...19M, 1992ASPC...26..416C, 2003ApJ...586..464B, 2013A&A...556A..36G}, and the index $b$ reflects the geometry of the stellar magnetic field \citep{1988ApJ...333..236K}. There are at least two solar twin binaries that display enhanced rotational velocities -- above $2\sigma$ from the expected -- and activity for their ages: HIP 19911 and HIP 67620; if we consider the revised age for HIP 103983 (Spina et al., in preparation), it can also be considered a fast rotator for its age.

Besides the enhanced rotational velocities and higher chromospheric activity (\citetalias{2014A&A...572A..48R}; \citealt{F16sub}), some of these binaries also display peculiar chemical abundances (\citetalias{2016A&A...590A..32T}; \citealt{2016A&A...587A..46D}). As pointed out by \citeauthor{2016A&A...587A..46D}, the ultra-depletion of beryllium, which is observed on HIP 64150, could be explained by the interaction of the main star with the progenitor of the white dwarf companion. In addition to HIP 64150, the confirmed binaries HIP 19911 and HIP 67620 also display clearly enhanced $\mathrm{[Y/Mg]}$ abundances \citepalias[see fig. 9 in][and the discussion in Section \ref{peculiar} of this paper]{2016A&A...590A..32T}.

One interesting aspect about stars with enhanced activity and rotation is that these characteristics were hypothesized to be the result of dynamo action from close-in giant planets \citep[see][and references therein]{2016ApJ...830L...7K}. In fact, some of our early results pointed out that the star HIP 68468, for which we inferred two exoplanets candidates \citepalias{2017A&A...597A..34M}, had an enhanced rotational velocity when compared to other solar twins of the same age. However, a more careful analysis showed that the enhancement was instead a contribution of macroturbulence. Another explanation for these enhancements is due to magnetic interactions with either a close-in or an eccentric giant planet \citep{2000ApJ...533L.151C}, but recent results obtained by, e.g., \citet{2016A&A...592A.143F} and \citet{2017MNRAS.465.2734M} show that they cannot explain such anomalies.

\defcitealias{2010exop.book...15M}{MC10}

In light of these intriguing results, we sought to better understand the nature of these solar twin multiple systems by studying their orbital parameters, and use them to search for explanations of the observed anomalies, especially stellar rotation. The orbital parameters can be estimated from the radial velocity data of the stars \citep[see, e.g.,][hereafter MC10]{2010exop.book...15M}, with the quality of the results depending strongly on the time coverage of the data.

\section{Radial velocities}

Our solar twins HARPS data\footnote{\footnotesize{Based on observations collected at the European Organisation for Astronomical Research in the Southern Hemisphere under ESO programs 188.C-0265, 183.D-0729, 292.C-5004, 077.C-0364, 072.C-0488, 092.C-0721, 093.C-0409, 183.C-0972, 192.C-0852, 091.C-0936, 089.C-0732, 091.C-0034, 076.C-0155, 185.D-0056, 074.C-0364, 075.C-0332, 089.C-0415, 60.A-9036, 075.C-0202, 192.C-0224, 090.C-0421 and 088.C-0323.}} are completely described in \citetalias{2014A&A...572A..48R}. Their radial velocities (RV) are automatically measured from the HARPS Data Reduction Software \textbf{(see Table \ref{HARPS_rvs})}, and the noise limit of the instrument generally remains around 1 m s$^{-1}$. In order to broaden the coverage of our RV data, we also obtained more datasets that were available in the literature and public databases, including the HARPS archival data for other programs.

The mass and other stellar parameters of the solar twins were estimated with high precision using the combined HARPS spectra and differential analysis owing to their similarity with the Sun \citep[see, e.g.,][]{2014ApJ...795...23B, 2010A&A...519A..87B, 2016A&A...589A..17Y}. The ages for the solar twins were obtained using Yonsei-Yale isochrones \citep{2001ApJS..136..417Y} and probability distribution functions as described in \citet{2013ApJ...764...78R} and in \citetalias{2014A&A...572A..48R}. The full description and discussion of the stellar parameters of the HARPS sample are going to be presented in a forthcoming publication (Spina et al., in preparation).

The additional radial velocities data obtained from online databases and the literature are summarized in Table \ref{add_rvs}. These are necessary to increase the time span of the observations to include as many orbital phases as possible at the cost of additional parameters to optimize for (see Section \ref{short}).

\begin{table}
\begin{center}
\caption{HARPS radial velocities for stars in the Solar Twin Planet Search program. This table is presented for guidance regarding the form and content of the online supplementary data, which is available in its entirety in machine-readable format.}
\begin{tabular}{ccc}
\toprule[\heavyrulewidth]
Julian Date & RV & $\sigma_{\mathrm{RV}}$ \\
(d) & (km s$^{-1}$) & (km s$^{-1}$) \\
\bottomrule[\heavyrulewidth]
\multicolumn{3}{c}{HIP 6407} \\
2455846.750257 & 6.816873 & 0.001020 \\
2455850.716200 & 6.811186 & 0.000997 \\
2455851.710847 & 6.806352 & 0.000886 \\
2455852.703837 & 6.801079 & 0.001140 \\
2456164.849296 & 6.204053 & 0.001059 \\
2456165.853256 & 6.203359 & 0.001049 \\
2456298.564449 & 5.954045 & 0.001029 \\
\hline
\multicolumn{3}{c}{HIP 14501} \\
2452937.683821 & 7.024814 & 0.000475 \\
2452940.727854 & 7.025289 & 0.000516 \\
2453001.575609 & 7.026322 & 0.000658 \\
2453946.941856 & 7.023707 & 0.000708 \\
\midrule[\heavyrulewidth]
\label{HARPS_rvs}
\end{tabular}
\end{center}
\end{table}

\begin{table*}
\begin{center}
\caption{Additional radial velocities from other programs and instruments.}
\begin{tabular}{lll}
\toprule[\heavyrulewidth]
Instrument/Program & References & Data available for (HIP numbers) \\
\bottomrule[\heavyrulewidth]
 CfA Digital Speedometers & \citet{2002AJ....124.1144L} & 65708 \\
 ELODIE\textsuperscript{a} & \citet{1996AandAS..119..373B, 2004PASP..116..693M} & 43297, 54582, 62039, 64150, 72043, 87769 \\
 SOPHIE\textsuperscript{b} & \citet{2011SPIE.8151E..15P} & 6407, 43297, 54582, 62039, 64150, 87769 \\
 Lick Planet Search & \citet{2014ApJS..210....5F} & 54582, 65708 \\
 AAT Planet Search & \citet{2015MNRAS.453.1439J} & 18844, 67620, 73241, 79578, 81746 \\
 Various & \citet{2016AJ....152...46W} & 67620 \\
 HIRES/Keck RV Survey\textsuperscript{c} & \citet{2017AJ....153..208B} & 14501, 19911, 62039, 64150, 72043, 103983 \\
\bottomrule[\heavyrulewidth]
\multicolumn{3}{l}{\textsuperscript{a}\footnotesize{Available at \url{http://atlas.obs-hp.fr/elodie/}.}}\\
\multicolumn{3}{l}{\textsuperscript{b}\footnotesize{Available at \url{http://www.obs-hp.fr/guide/sophie/data_products.shtml}.}}\\
\multicolumn{3}{l}{\textsuperscript{c}\footnotesize{Available at \url{http://home.dtm.ciw.edu/ebps/data/}.}}\\
\label{add_rvs}
\end{tabular}
\end{center}
\end{table*}

\section{Methods}

The variation of radial velocities of a star in binary or multiple system stems from the gravitational interaction between the observed star and its companions. For systems with stellar or substellar masses, the variation of radial velocities can be completely explained by the Keplerian laws of planetary motion. For the sake of consistency, we will use here the definitions of orbital parameters as presented in \citetalias{2010exop.book...15M}.

To completely characterize the orbital motion of a binary system from the measured radial velocities of the main star, we need to obtain the following parameters: the semi-amplitude of the radial velocities $K$, the orbital period $T$, the time of periastron passage $t_0$, the argument of periapse $\omega$ and the eccentricity $e$ of the orbit. In order to estimate the minimum mass $m \sin{i}$ of the companion and the semi-major axis $a$ of the orbit, we need to know the mass $M$ of the main star.

Due to their non-negative nature, the parameters $K$ and $T$ are usually estimated in logarithmic scale in order to eliminate the use of search bounds. Additionally, for orbits that are approximately circular, the value of $\omega$ may become poorly defined. In these cases, a change of parametrization may be necessary to better constrain them. \citet{2013PASP..125...83E}, for instance, suggest using $\sqrt{e} \cos{\omega}$ and $\sqrt{e} \sin{\omega}$ (which we refer to as the EXOFAST parametrization) instead of $\omega$ and $e$ to circumvent this problem, which also can help improve convergence time.

One issue that affects the radial velocities method is the contamination by stellar activity \citep[see, e.g.,][]{2016MNRAS.457.3637H}. This activity distorts the spectral lines \citep{2005oasp.book.....G}, which in turn produces artificial RV variations that can mimic the presence of a massive companion orbiting the star. More active stars are expected to have RV variations with larger amplitudes and a shorter activity cycle period \citep{2011arXiv1107.5325L}. For most binaries in our sample, the contamination by activity in the estimation of orbital parameters is negligible; the cases where this is not applicable are discussed in detail.

\subsection{Binaries with well-sampled orbits}\label{short}

For binaries with orbital periods $T \lesssim 15$ yr, usually there are enough RV data measured to observe a complete phase. In these cases, the natural logarithm of the likelihood of observing radial velocities $\mathbf{y}$ on a specific instrument, given the Julian dates $\mathbf{x}$ of the observations, their uncertainties $\mathbf{\sigma}$ and the orbital parameters $\mathbf{p}_{\mathrm{orb}}$ is defined as:

\begin{equation}
 \ln{p \left(\mathbf{y} \mid \mathbf{x},\mathbf{\sigma}, \mathbf{p}_{\mathrm{orb}} \right)} = -\frac{1}{2} \sum_n \left[ \frac{ \left(y_\mathrm{n} - y_{\mathrm{model}} \right)^2}{\sigma_\mathrm{n}^2} + \ln{ \left(2 \pi \sigma_\mathrm{n}^2 \right)} \right] \mathrm{,}
\label{likelihood}
\end{equation}where $y_\mathrm{n}$ are the RV datapoints, $y_{\mathrm{model}}$ are the model RV points for a given set of orbital parameters, and $\sigma_\mathrm{n}$ are the RV point-by-point uncertainties. The RV models are computed from Eq. 65 in \citetalias{2010exop.book...15M}:

\begin{equation}
 v_\mathrm{r} = \gamma + K (\cos{(\omega + f)} + e \cos{\omega}) \mathrm{,}
\end{equation}where $f$ is the true anomaly, and $\gamma$ is the systemic velocity (usually including the instrumental offset). The true anomaly depends on $e$ and the eccentric anomaly $E$:

\begin{equation}
 \cos{f} = \frac{(1 - e^2)}{1 - e \cos{E}} - 1 \mathrm{;}
\end{equation}the eccentric anomaly, in turn, depends on $T$, $t_0$ and time $t$ in the form of the so called Kepler's equation:

\begin{equation}
 E - e \sin{E} = \frac{2 \pi}{T} \left(t - t_0 \right) \mathrm{.}
\end{equation}
Eq.~\ref{likelihood} is minimized using the Nelder-Mead algorithm implementation from \texttt{lmfit} \citep[][version 0.9.5]{2016ascl.soft06014N} to obtain the best-fit orbital parameters to the observed data. Because different instruments have different instrumental offsets, the use of additional RV data from other programs require the estimation of an extra value of $\gamma$ for each instrument.

The uncertainties of the orbital parameters are estimated using \texttt{emcee}, an implementation of the Affine Invariant Markov chain Monte Carlo Ensemble sampler \citep[][version 2.2.1]{2013PASP..125..306F} using flat priors for all parameters in both \citetalias{2010exop.book...15M} and EXOFAST parametrizations. These routines were implemented in the Python package \texttt{radial}\footnote{\footnotesize{Available at \url{https://github.com/RogueAstro/radial}.}}, which is openly available online. The uncertainties in $m \sin{i}$ and $a$ quoted in our results already take into account the uncertainties in the stellar masses of the solar twins.

\subsection{Binaries with partial orbits}\label{methods_long_period}

For the binary systems with long periods (typically 20 years or more), it is possible that the time span of the observations does not allow for a full coverage of at least one phase of the orbital motion. In these cases, the estimation of the orbital parameters renders a number of possible solutions, which precludes us from firmly constraining the configuration of the system. Nevertheless, RV data containing a curvature or one inflection allows us to place lower limits on $K$, $T$ and $m \sin{i}$, whilst leaving $e$ and $\omega$ completely unconstrained. When the RV data are limited but comprise two inflections, it may be possible to use the methods from Section \ref{short} to constrain the orbital parameters, albeit with large uncertainties.

For stars with very large orbital periods ($T \gtrsim 100$ yr), the variation of radial velocities may be present in the form of a simple linear trend. In these cases, it is still possible to obtain an estimate of the mass of the companion -- a valuable piece of information about it: \citet{1999PASP..111..169T} describes a statistical approach to extract the sub-stellar companion mass when the only information available from radial velocities is the inclination of the linear trend, provided information about the angular separation of the system is also available. In this approach, we need to adopt reasonable prior probability density functions (PDF) for the eccentricity $e$, the longitude of periastron $\varpi$, phase $\phi$ and the inclination $i$ of the orbital plane. As in \citeauthor{1999PASP..111..169T}, we adopt the following PDFs: $p(i) = \sin{i}$, $p(e) = 2e$ and flat distributions for $\varpi$ and $\phi$.

We sample the PDFs using \texttt{emcee}, 20 walkers and 10000 steps; the first 500 burn-in steps are discarded. From these samples, we compute the corresponding companion masses and its posterior distribution. This posterior usually displays a very strong peak and long tails towards low and high masses which can be attributed to highly unlikely orbital parameters (see Fig. \ref{64150_pdf} for an example). In our results, we consider that the best estimates for the companion masses are the central bin of the highest peak of the distribution in a histogram with log-space bin widths of about 0.145 dex(M$_\odot$).

When no adaptive optics (AO) imaging data are available for the stars with a linear trend in their RVs, the most conservative approach is to provide the minimum mass for the putative companion. In the case of a linear trend, the lowest mass is produced when $e = 0.5$, $\omega = \pi/2$ and $\sin{i} = 1$ \citep{2015ApJ...800...22F}, yielding

\begin{equation}\label{feng_eq}
  m_{\mathrm{min}} \approx \left( 0.0164\ \mathrm{M}_{\mathrm{Jup}}\right) \left( \frac{\tau}{\mathrm{yr}} \right)^{4/3} \left| \frac{dv/dt}{\mathrm{m\ s}^{-1}\ \mathrm{yr}^{-1}} \right| \left( \frac{M}{\mathrm{M}_\odot} \right)^{2/3} \mathrm{,}
\end{equation}where $\tau$ is 1.25 multiplied by the time span of the radial velocities and $dv/dt$ is the inclination of the linear trend.

\section{Results}

\begin{figure*}
\centering
\begin{tabular}{cc}
\includegraphics[width=0.47\textwidth]{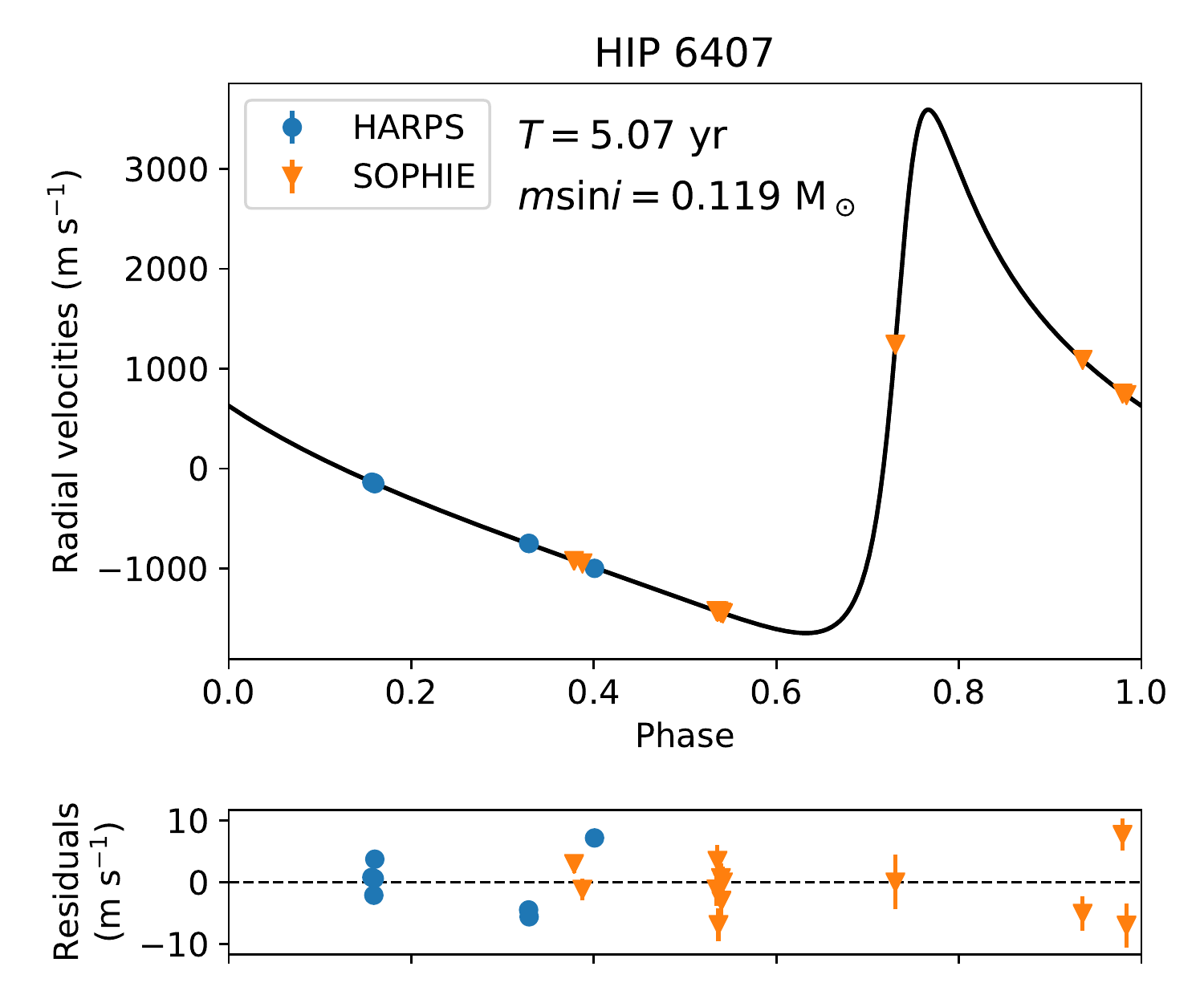} & \includegraphics[width=0.47\textwidth]{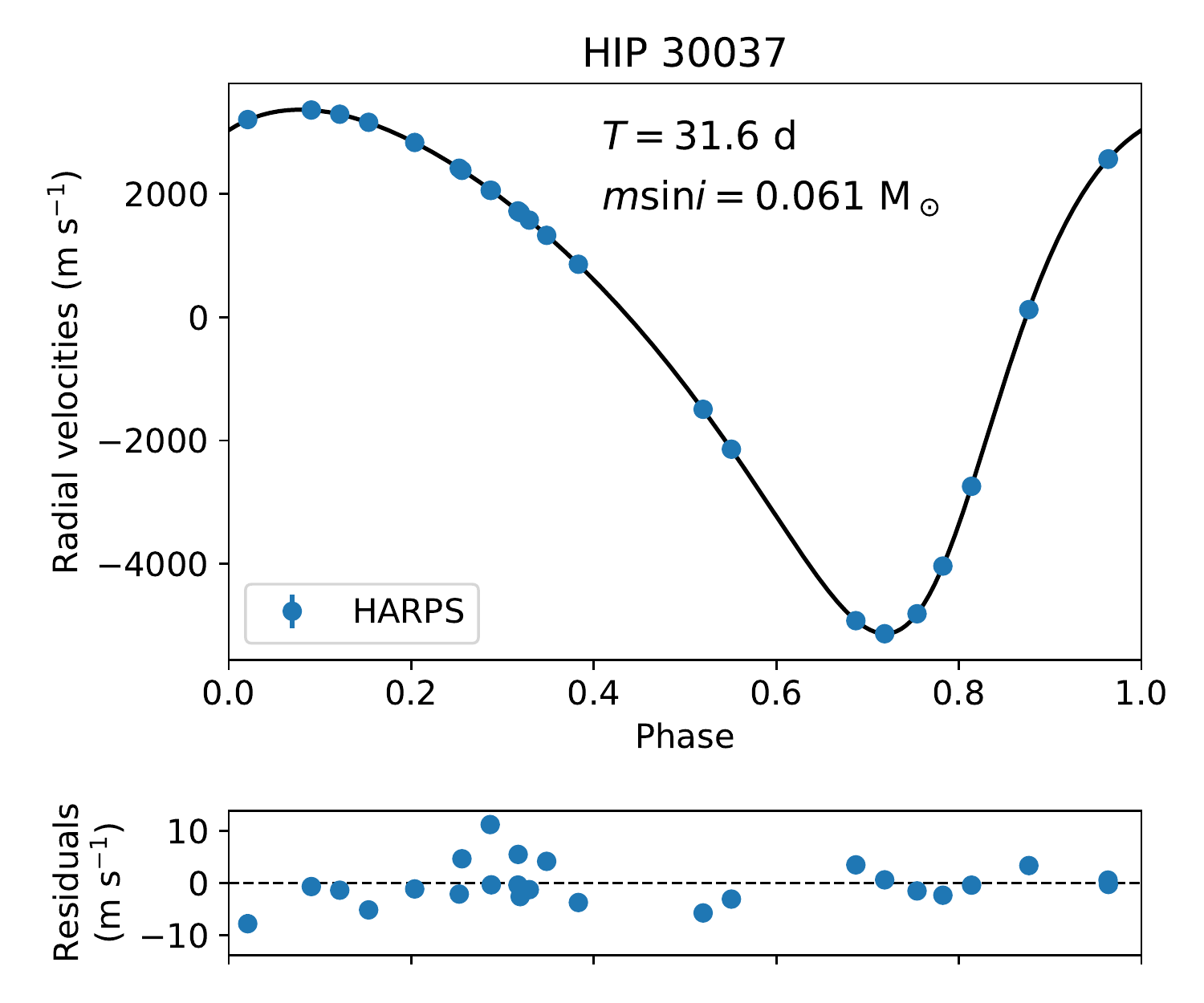} \\
\end{tabular}
\caption{The radial velocities and orbital solutions of the newly discovered companions for solar twins with short orbital periods.}
\label{orbits_new}
\end{figure*}

\begin{figure*}
\centering
\begin{tabular}{cc}
\includegraphics[width=0.45\textwidth]{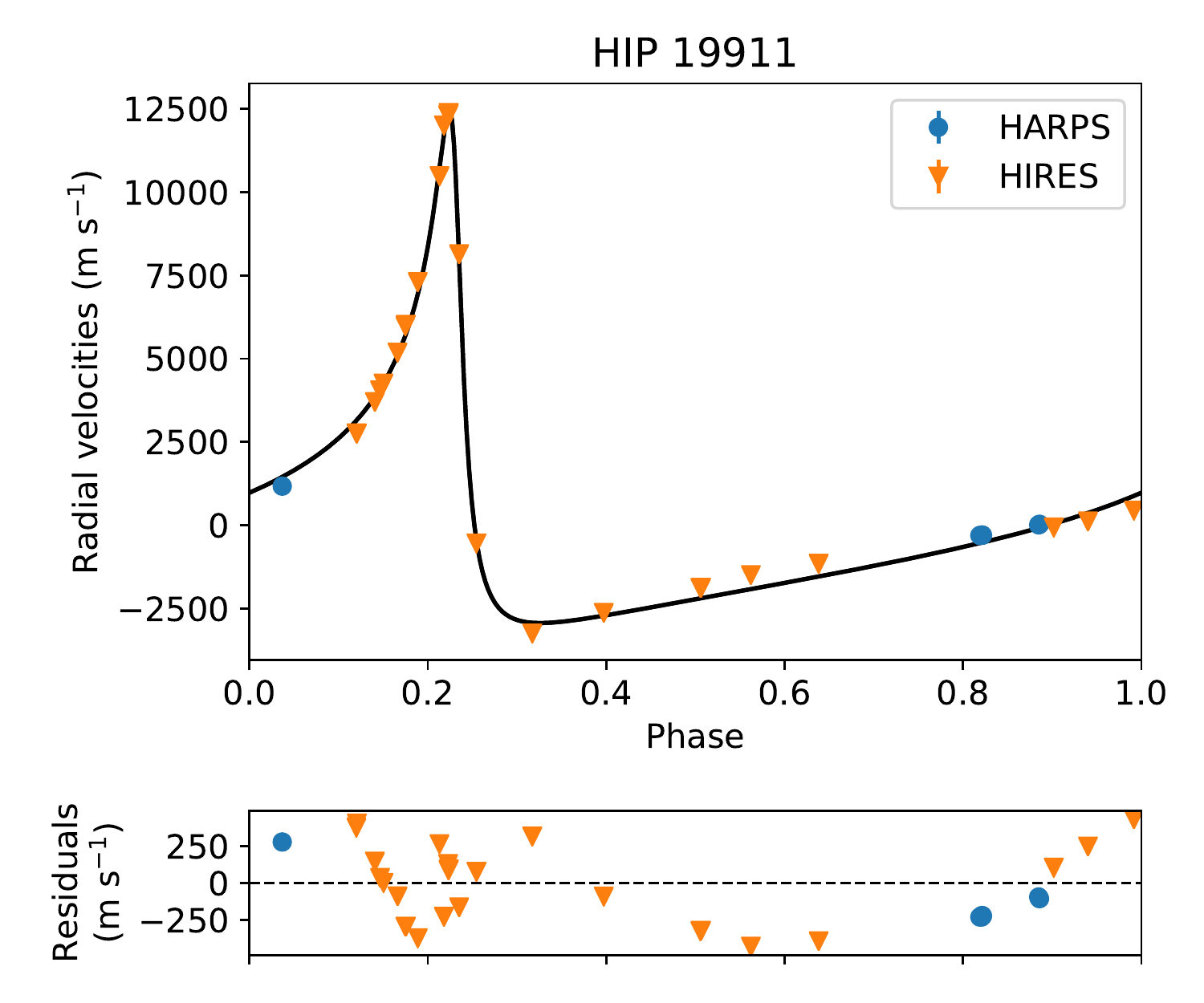} & \includegraphics[width=0.45\textwidth]{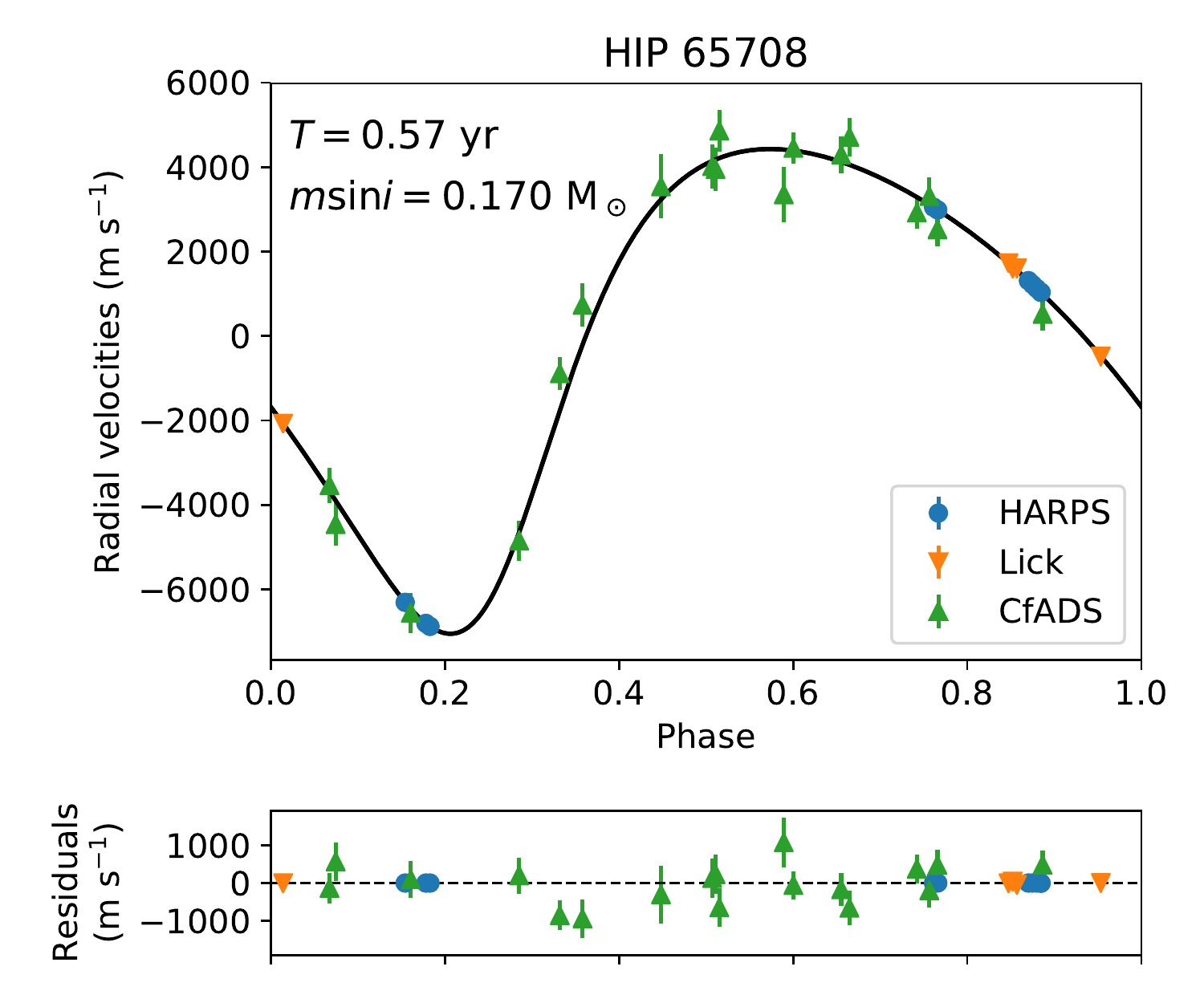} \\
\includegraphics[width=0.45\textwidth]{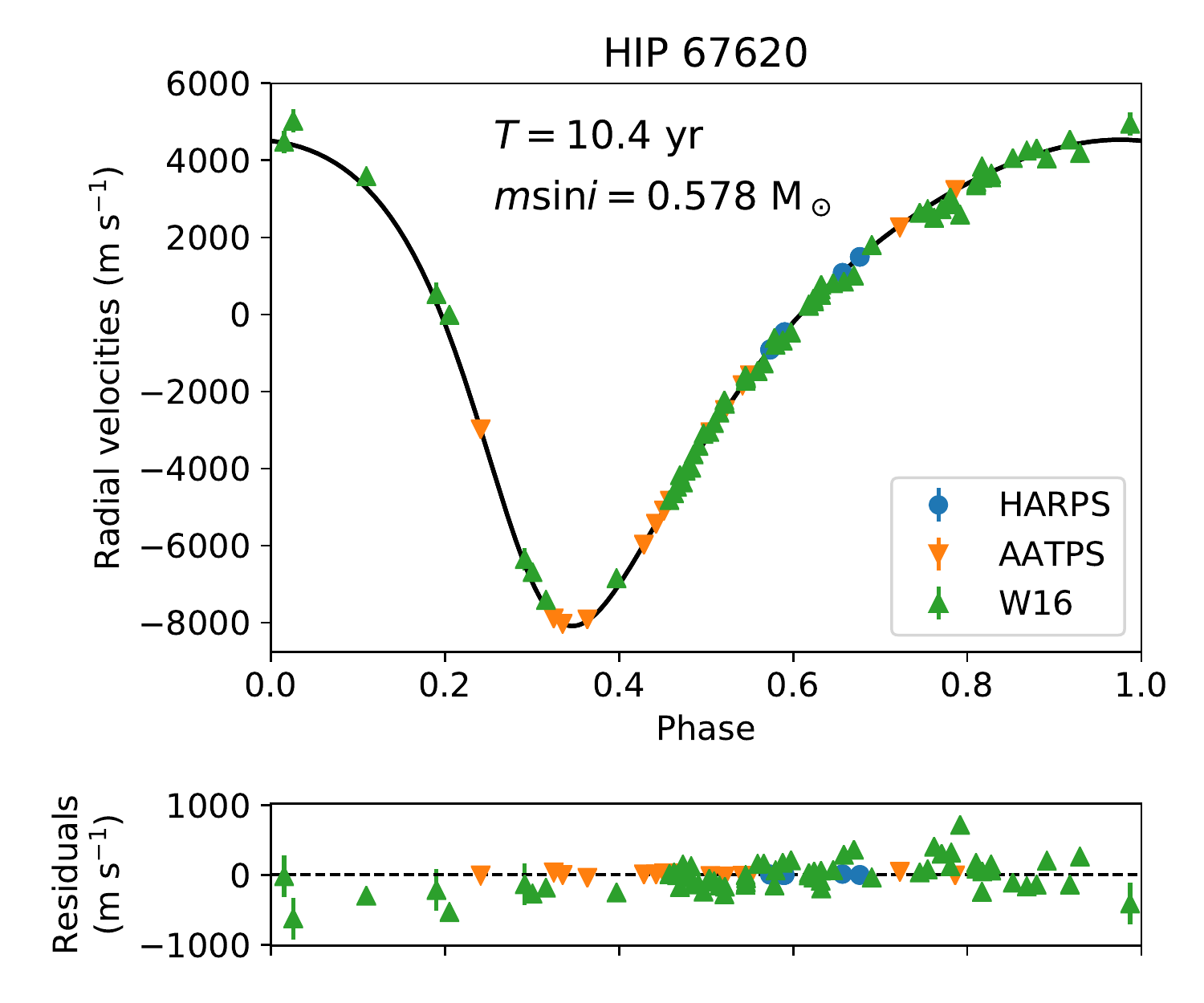} & \includegraphics[width=0.45\textwidth]{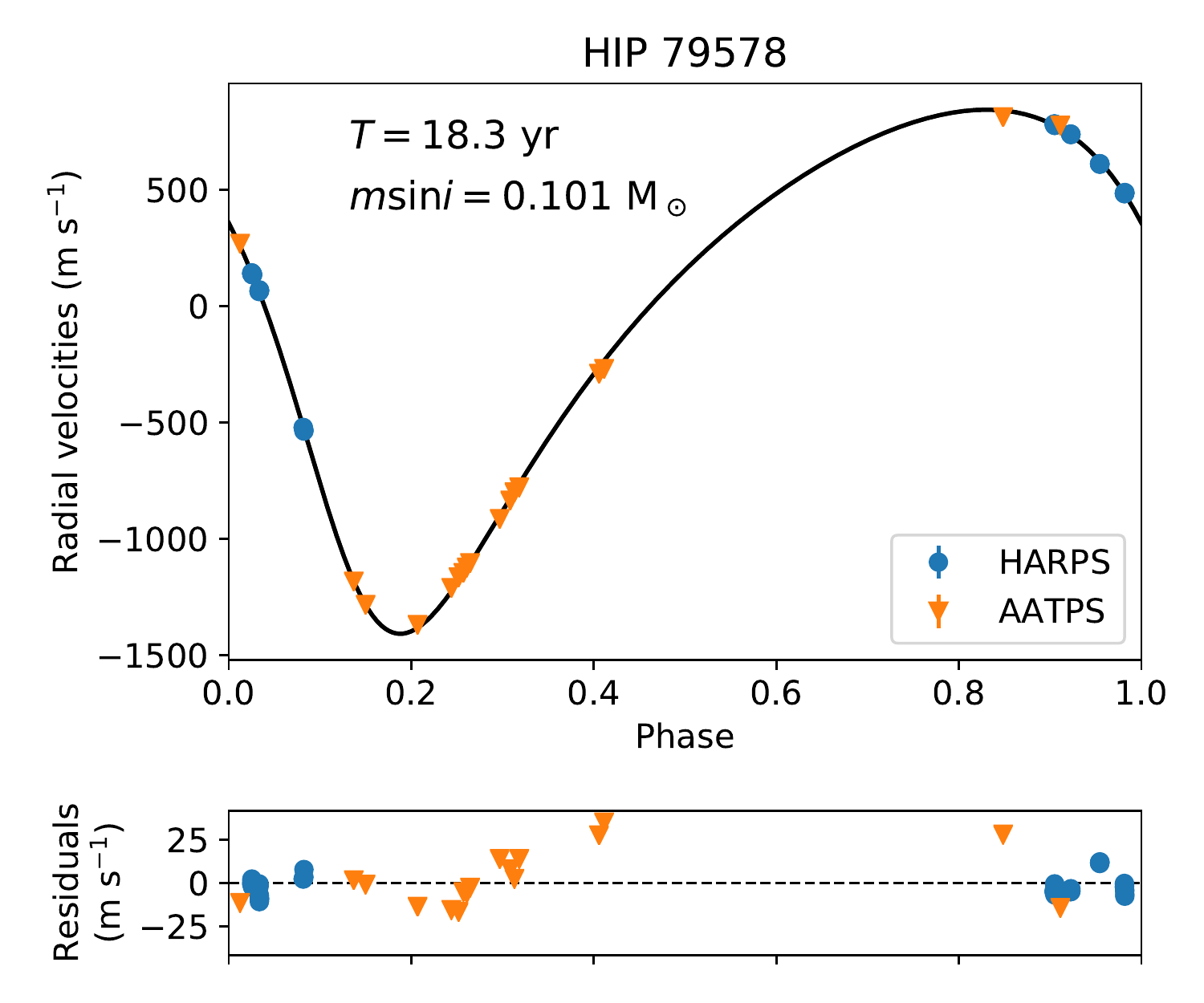} \\
\includegraphics[width=0.45\textwidth]{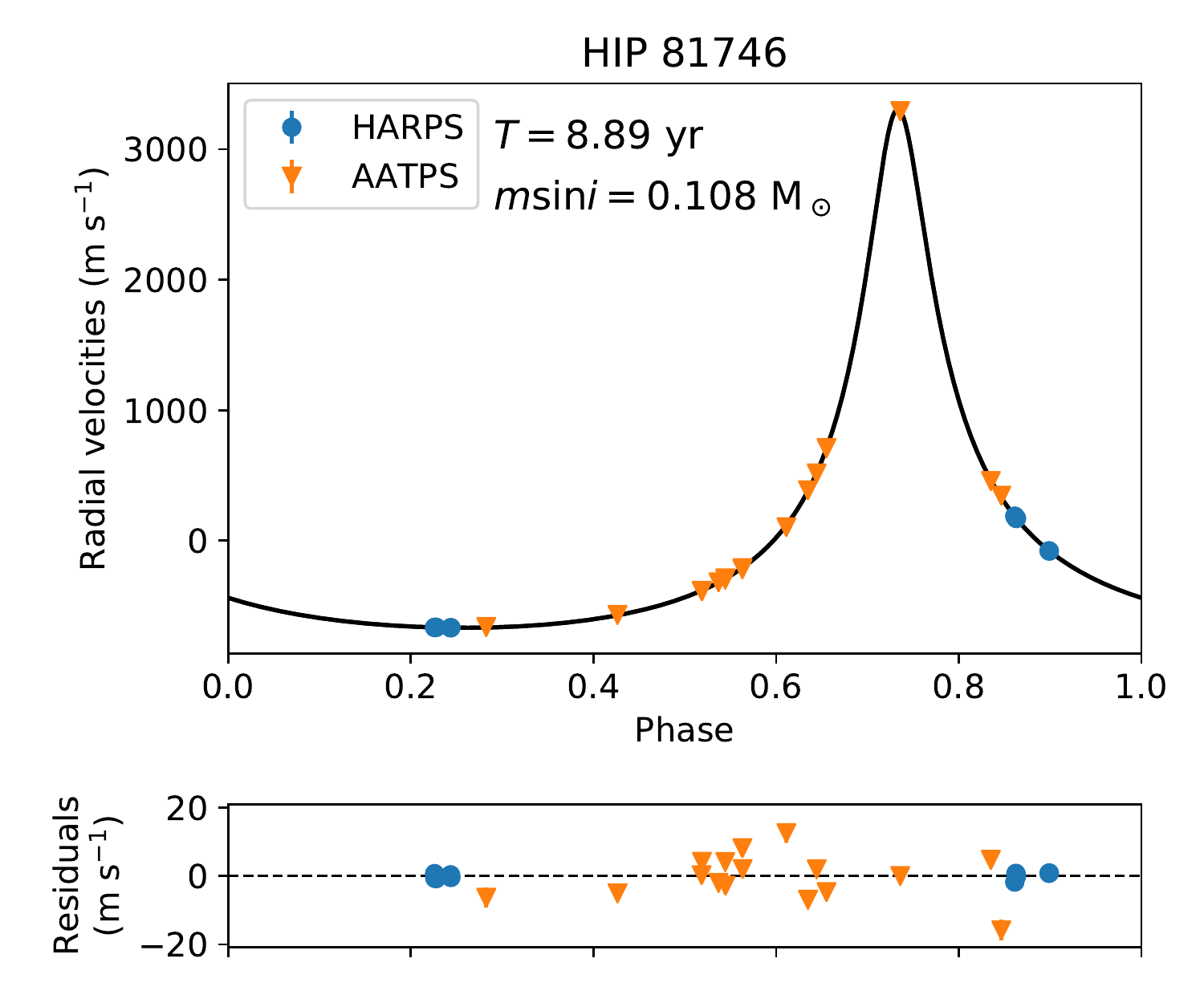} & \includegraphics[width=0.45\textwidth]{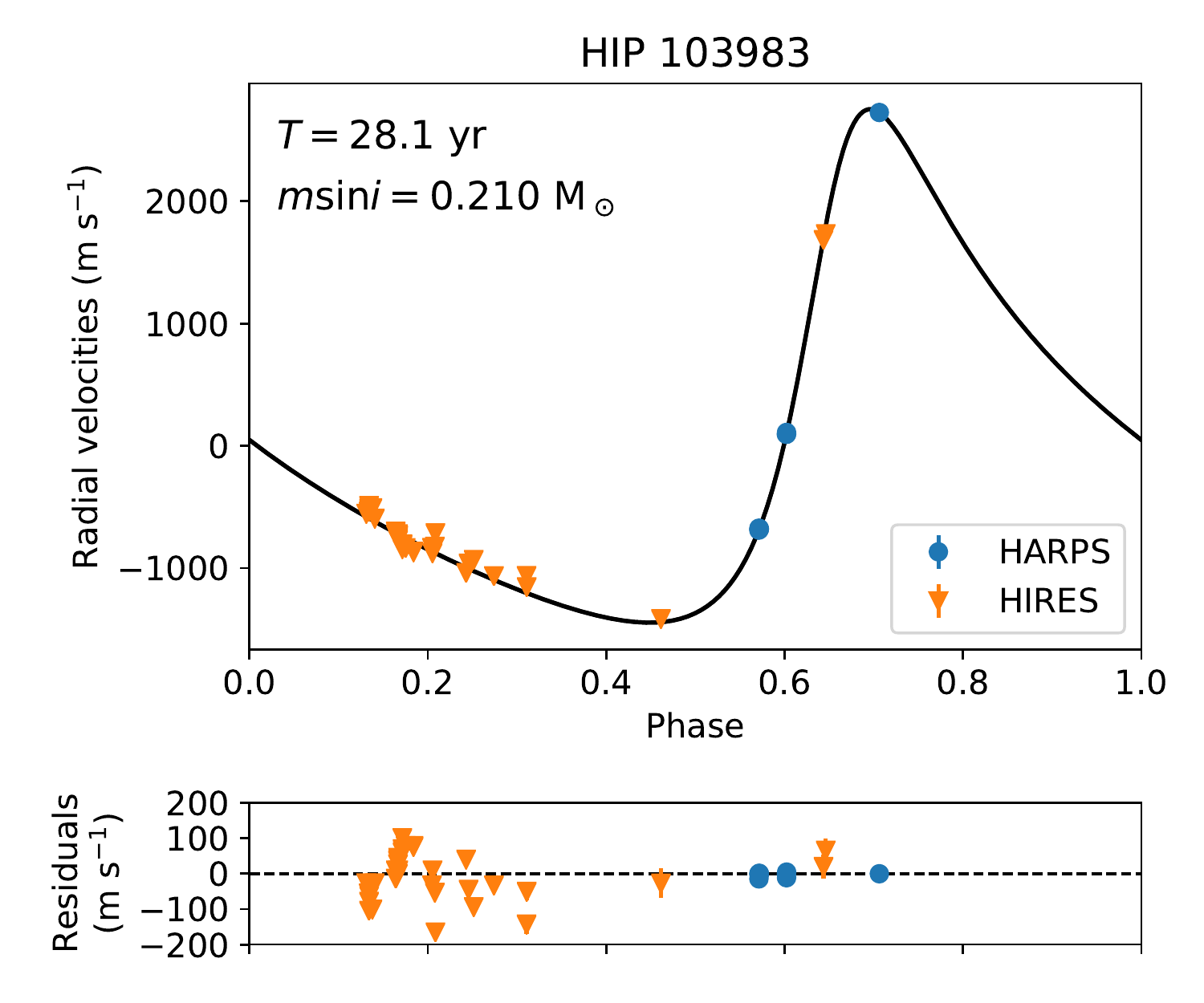} \\
\end{tabular}
\caption{The radial velocities and updated orbital solutions of the known solar twin binaries with at least one complete or near-complete orbital phase. The acronyms CfADS, W16 and AATPS correspond to data from, respectively, the Center for Astrophysics Digital Speedometers, \citet{2016AJ....152...46W} and the Anglo-Australian Telescope Planet Search program.}
\label{orbits_updated}
\end{figure*}

We discovered new, short-period companions for the stars HIP 6407 and HIP 30037 (see Fig. \ref{orbits_new}) and new long-period companions for HIP 54582 and HIP 62039, and updated or reproduced the parameters of several other known binaries that were observed in our program (see Figs. \ref{orbits_updated} and \ref{long_period_rvs}). We briefly discuss below each star, pointing out the most interesting results, inconsistencies and questions that are still open about each of them. The orbital parameters of the binaries with well-sampled orbits in their RV data are presented in Table \ref{short_params} and the systems with partial orbits are reported in Tables \ref{curvature_results} and \ref{linear_trend_results}.

\subsection{Withdrawn binary candidates}

In \citetalias{2016A&A...592A.156D}, we showed that HIP 43297 had a rotational velocity $v \sin{i}$ higher than expected for its age. Moreover, its radial velocities had variations that hinted for one or more companions orbiting it. We carefully analyzed the RVs and concluded that the periodic ($T = 3.8$ yr) signal observed is highly correlated (Pearson $R = 0.893$) with the activity \textit{S}-index of the star \citep{F16sub}. In addition, we tentatively fitted a linear trend to the combined RVs from HARPS, ELODIE and SOPHIE, and obtained an inclination of $4.53 \pm 0.04$ m s$^{-1}$ yr$^{-1}$, but further monitoring of the system is required to infer the presence of a long-period spectroscopic companion. The revised stellar age for HIP 43297 yields $1.85 \pm 0.50$ Gyr (Spina et al., in preparation), which explains the high rotational velocity and activity.

The solar twin HIP 64673 displays significant fluctuations in its radial velocities, but they do not correlate with its activity index; the data covers approximately 5 years of RV monitoring and displays an amplitude $> 20$ m s$^{-1}$. If confirmed to be caused by massive companions, the RV variations of both HIP 43297 and HIP 64673 suggest substellar masses for the most likely orbital configurations. These stars are, thus, removed from the binaries sample of the Solar Twin Planet Search program.

\begin{table*}
\begin{center}
\caption{Orbital parameters of the binaries with complete or near-complete orbital phases in their RV data.}
\begin{tabular}{llccccccc}
\toprule[\heavyrulewidth]
\multirow{2}{*}{HIP} & \multirow{2}{*}{HD} & $K$ & $T$ & $t_0$ & $\omega$ & $e$ & $m\sin{i}$ & $a$ \\
& & (km s$^{-1}$) & (days) & (JD-$2.45E6$ days) & ($^o$) & & (M$_\odot$) & (AU) \\
\bottomrule[\heavyrulewidth]
\multirow{2}{*}{6407\textsuperscript{$\dag$}} & \multirow{2}{*}{8291} & $2.614$ & $1852.3$ & $5076.7$ & $-57.1$ & $0.682$ & $0.119$ & $3.070$ \\
& & $\pm 0.084$ & $^{+3.3}_{-3.1}$ & $^{+1.1}_{-1.3}$ & $\pm 0.9$ & $^{+0.009}_{-0.010}$ & $\pm 0.002$ & $\pm 0.005$ \\
\hline
\multirow{2}{*}{19911} & \multirow{2}{*}{26990} & $7.707$ & $2074.15$ & $4627.4$ & $40.92$ & $0.8188$ & $0.313$ & $6.16$ \\
& & $\pm 0.007$ & $\pm 0.09$ & $\pm 0.1$ & $\pm 0.06$ & $\pm 0.0003$ & $\pm 0.002$ & $\pm 0.02$ \\
\hline
\multirow{2}{*}{30037} & \multirow{2}{*}{45021} & $4.246$ & $31.61112$ & $5999.413$ & $-133.60$ & $0.30205$ & $0.0610$ & $0.1971$ \\
& & $\pm 0.003$ & $\pm 0.00006$ & $\pm 0.001$ & $\pm 0.02$ & $\pm 0.00008$ & $\pm 0.0002$ & $\pm 0.0003$ \\
\hline
\multirow{2}{*}{65708} & \multirow{2}{*}{117126} & $5.754$ & $207.273$ & $-3675.8$ & $-137.9$ & $0.311$ & $0.170$ & $0.851$ \\
& & $\pm 0.007$ & $\pm 0.004$ & $\pm 0.3$ & $\pm 0.2$ & $\pm 0.002$ & $\pm 0.001$ & $\pm 0.001$ \\
\hline
\multirow{2}{*}{67620} & \multirow{2}{*}{120690} & $6.311$ & $3803.3$ & $4945.7$ & $145.10$ & $0.3428$ & $0.578$ & $5.50$ \\
& & $\pm 0.002$ & $\pm 0.4$ & $\pm 0.7$ & $\pm 0.07$ & $\pm 0.0002$ & $\pm 0.002$ & $\pm 0.01$ \\
\hline
\multirow{2}{*}{79578} & \multirow{2}{*}{145825} & $1.125$ & $6681.8$ & $879.2$ & $138.9$ & $0.3322$ & $0.1014$ & $7.216$ \\
& & $\pm 0.007$ & $\pm 1.5$ & $\pm 1.9$ & $\pm 0.1$ & $\pm 0.0003$ & $\pm 0.0002$ & $\pm 0.007$ \\
\hline
\multirow{2}{*}{81746} & \multirow{2}{*}{150248} & $1.987$ & $3246.5$ & $5623.8$ & $-2.49$ & $0.6644$ & $0.1079$ & $4.387$ \\
& & $\pm 0.001$ & $\pm 0.7$ & $\pm 0.6$ & $\pm 0.06$ & $\pm 0.0005$ & $\pm 0.0002$ & $\pm 0.003$ \\
\hline
\multirow{2}{*}{103983} & \multirow{2}{*}{200565} & $2.100$ & $10278$ & $6659$ & $-51.6$ & $0.50$ & $0.210$ & $9.8$ \\
& & $\pm 0.018$ & $^{+274}_{-247}$ & $\pm 11$ & $\pm 1.5$ & $\pm 0.01$ & $\pm 0.005$ & $\pm 0.2$ \\
\midrule[\heavyrulewidth]
\multicolumn{9}{l}{\textsuperscript{$\dag$}\footnotesize{Triple or higher-order system. The orbital parameters correspond to the closer-in companion.}}\\
\end{tabular}
\label{short_params}
\end{center}
\end{table*}

\subsection{Solar twins with new companions}

\textbf{HIP 6407:} This is a known binary system located 58 pc away from the solar system \citep{2007A&A...474..653V}, possessing a very low-mass (0.073 M$_\odot$) L2-type companion separated by $44.8\arcsec$ (2222 AU), as reported by \citet[][and references therein]{2015ApJ...802...37B}. In this study, we report the detection of a new close-in low-mass companion with $m\sin{i} = 0.12$ M$_\odot$ on a very eccentric orbit ($e = 0.67$) with $a = 3$ AU and an orbital period of approximately 5 years. As expected, the long-period companion does not appear in the RV data as a linear trend.

\textbf{HIP 30037:} The most compact binary system in our sample, hosting a brown dwarf companion orbiting the main star with a period of 31 days. The high precision of its parameters owes to the wide time span of observations, which covered several orbits. This is one of the first detections of a close-in brown dwarf orbiting a confirmed solar twin\footnote{\footnotesize{There are at least 4 solar twin candidates with a close-in brown dwarf companion listed in table A.1 in \citet{2016A&A...588A.144W}.}}. HIP 30037 is a very quiet star, displaying no excessive jitter noise in its radial velocities. We ran stellar evolution models with \texttt{MESA}\footnote{\footnotesize{Modules for Experiments in Stellar Astrophysics, available at \url{http://mesa.sourceforge.net}}} \citep{2011ApJS..192....3P, 2015ApJS..220...15P} to test the hypothesis of the influence of tidal acceleration caused by the companion on a tight orbit, and found that, for the mass and period of the companion, we should expect no influence in the rotational velocity.

\textbf{HIP 54582:} \textit{RV Curvature only}. There are no reports of binarity in the literature. The slight curvature in the RVs of this star is only visible when we combine the HARPS data and the Lick Planet Search archival data. Owing to the absence of an inflection point, the orbital parameters of this system are highly unconstrained. We found that an orbit with $e \approx 0.2$ produces the least massive companion and shortest orbital period ($m \sin{i} = 0.03$ M$_\odot$ and $T = 102$ yr).

\textbf{HIP 62039:} \textit{Linear trend}. There are no reports of visually detected close-in ($\rho < 2\arcsec$) companions around it. This can be attributed to: i) low luminosity companion, which is possible if it is a white dwarf or a giant planet, and ii) unfavorable longitude of periapse during the observation windows. By using Eq. \ref{feng_eq}, we estimate that the minimum mass of the companion is 19 M$_{\mathrm{Jup}}$.

\subsection{The peculiar binaries}\label{peculiar}

\subsubsection{HIP 19911}\label{19911_results}

This is one of the main outlier stars in the overall sample of solar twins in regards to its rotation and activity, which are visibly enhanced for both the previous and revised ages (\citetalias{2016A&A...592A.156D}; \citealt{F16sub}; Spina et al., in preparation). For the estimation of orbital parameters reported below, we used only the LCES HIRES/Keck radial velocities, because there are too few HARPS data points to justify the introduction of an extra source of uncertainties (the HARPS points are, however, plotted in Fig. \ref{orbits_updated} for reference). When using the HARPS data, although the solution changes slightly, our conclusions about the system remain the same.

The orbital solution of HIP 19911 renders a 0.31 M$_\odot$ companion in a highly eccentric orbit ($e = 0.82$, the highest in our sample), with period $T = 5.7$ yr. Visual scrutiny reveals what seems to be another signal with large amplitude  in the residuals of this fit ($> 250$ m s$^{-1}$, see Fig. \ref{orbits_updated}); the periodogram of the residuals shows a very clear peak near the orbital period of the stellar companion.

\begin{figure*}
\centering
\begin{tabular}{cc}
\includegraphics[width=0.47\textwidth]{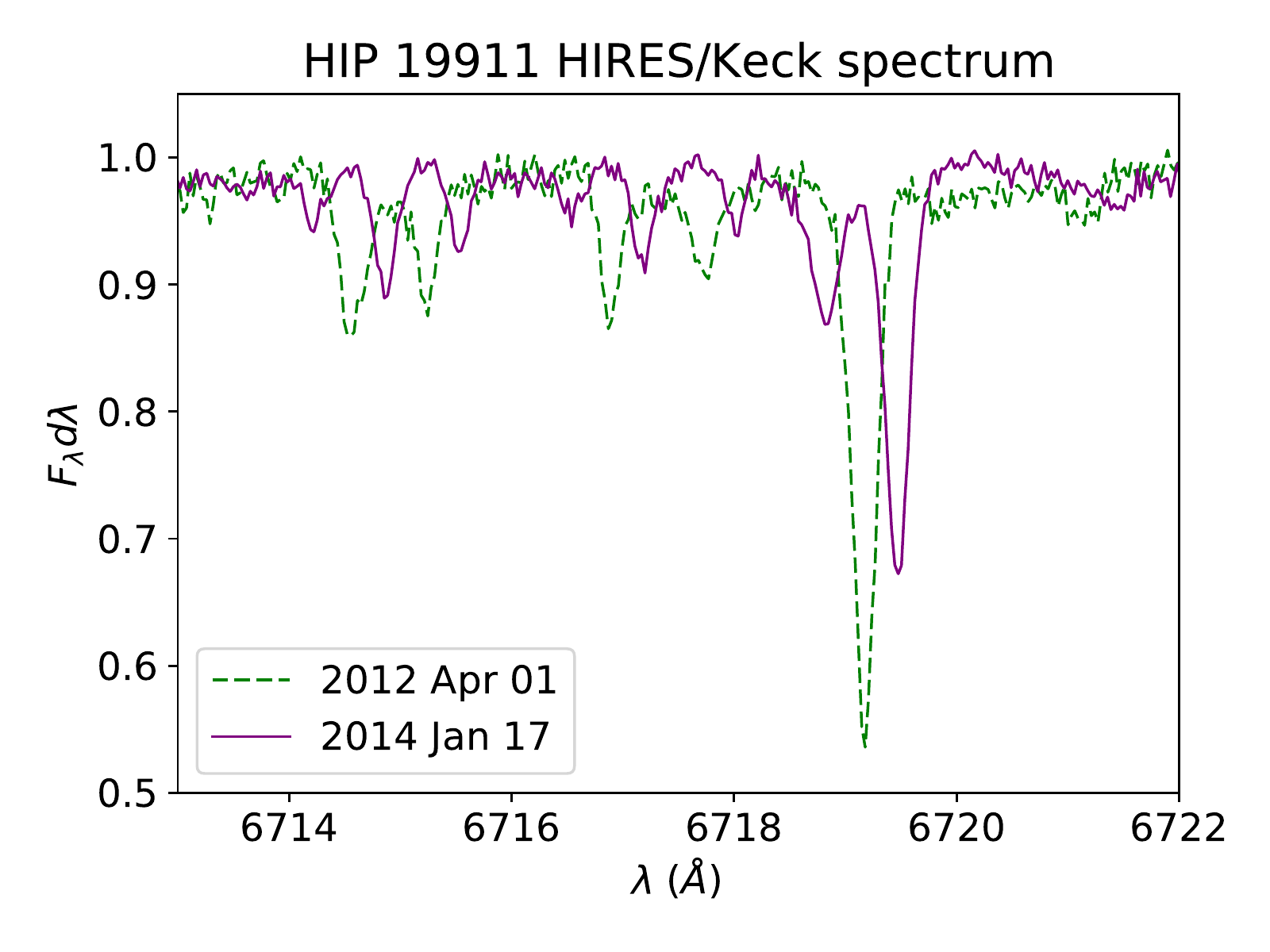} & \includegraphics[width=0.47\textwidth]{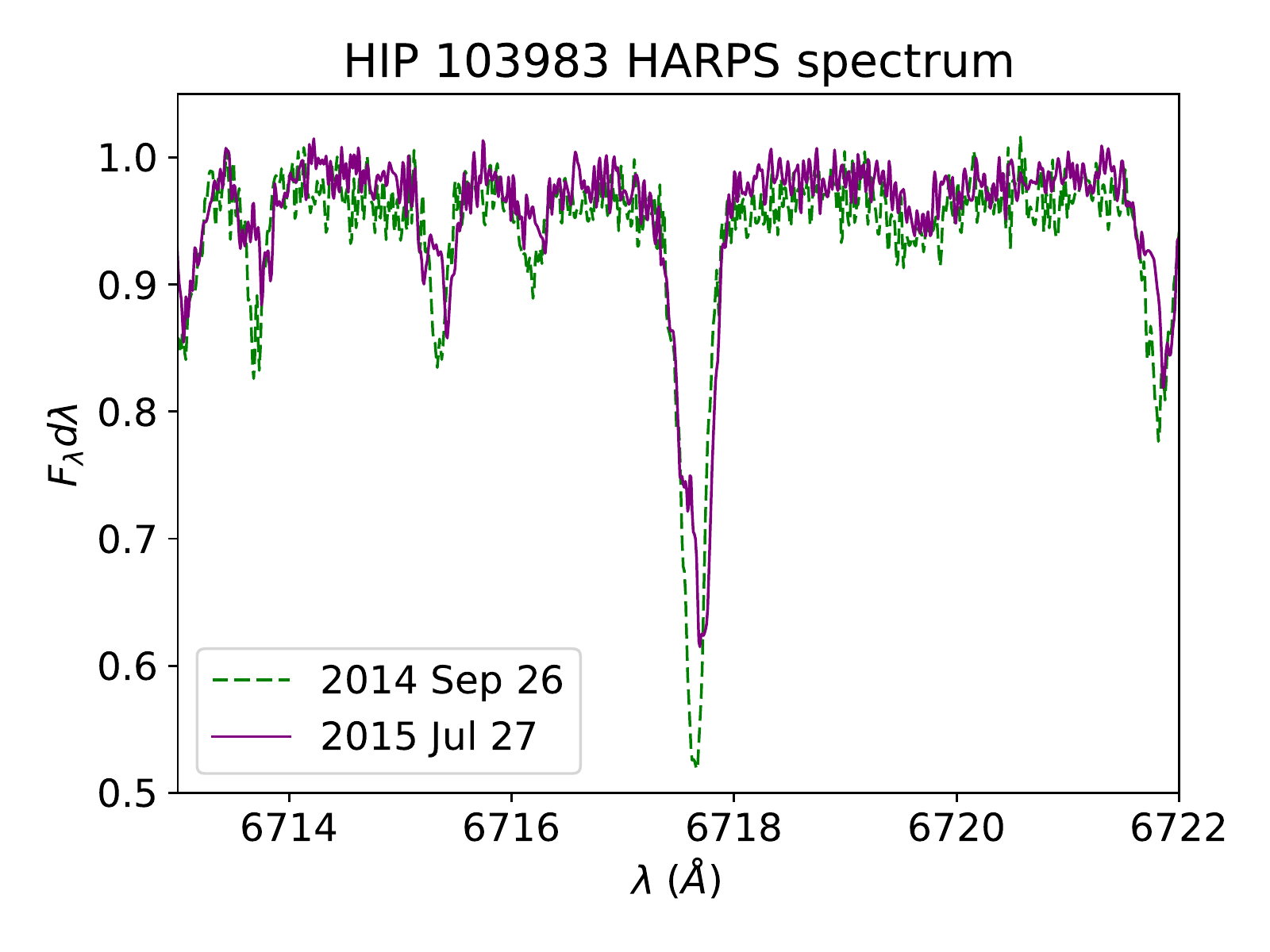} \\
\end{tabular}
\caption{The spectrum of HIP 19911 and HIP 103983 are contaminated by bright companions. The double-lined nature of these stars is only clear only in the maximum RV separation, which are the solid purple curves. The dashed green curves show spectra during minimum RV separation where there is no obvious contamination. The spectra in this figure are not Doppler-corrected for their systemic velocities or instrumental RV offset.}
\label{SBII}
\end{figure*}

The cross-correlation function (CCF) plots for the HARPS spectra of HIP 19911 display a significant asymmetry -- longer tail in the blue side -- for the observations between October 2011 and February 2012, which suggests that the companion is contaminating the spectra. Upon visual inspection of the archival HIRES spectra\footnote{\footnotesize{Available at \url{http://nexsci.caltech.edu/archives/koa/}.}} taken on 17 January 2014, which is when we expect the largest RV difference between the main star and its companion, we saw a clear contamination of the spectrum by the companion (see Fig. \ref{SBII}). This contamination could explain the large residuals of the orbital solution, as it introduces noise to the measured radial velocities. The double-lines also explain the inferred high rotational velocity of HIP 19911, since they introduce extra broadening to the spectral lines used to measure rotation. The presence of a bright companion may also affect estimates of chemical abundances, which elucidates the yttrium abundance anomaly \citepalias{2016A&A...590A..32T}. The double-lined nature of this system is not observed on the HARPS spectra due to an unfavorable observation window.

Even at the largest RV separation, we did not detect the Li I line at 6707.75 \AA\ in the HIRES spectrum of the companion. This is expected because M dwarf stars have deeper convection zones, which means they deplete lithium much faster than Sun-like stars. This leads us to conclude that estimates of Li abundance on solar twin binaries using this line do not suffer from strong contamination by their companions; consequently, age estimates with lithium abundances may be more reliable for such binaries than isochronal or gyro ages.

\defcitealias{2014AJ....147...86T}{T14}

Another observation conundrum for this system is that \citet{2015ApJ...799....4R}, using AO imaging without a coronagraph, reports the detection of a visual companion with orbital period $\sim 12.4$ yr (roughly twice the one we estimated), a lower eccentricity ($e = 0.1677$) and similar semi-major axis of the orbit ($a = 6.17$ AU, if we consider a distance of 30.6 pc). Moreover, \citet[][hereafter T14]{2014AJ....147...86T} reports that this visual companion has $m = 0.85$ M$_\odot$. The most likely explanation is that the observations of \citeauthor{2015ApJ...799....4R} did in fact detect the spectroscopic companion, but the coarse timing of the observations produced a larger period; the lower eccentricity could be explained by a strong covariance between $e$ and the inclination $i$. If $i$ is lower, that means the mass of the companion is significantly higher than $m \sin{i} = 0.316$ M$_\odot$, and that would explain the value obtained by \citetalias{2014AJ....147...86T}. A companion with a mass as large as 0.85 M$_\odot$ would likely pollute the spectra of HIP 19911, which agrees with our observation that this is a SB II system. If confirmed, this prominent $\sim$0.85 M$_\odot$ red dwarf companion could explain the observed activity levels for HIP 19911, since red dwarf stars are expected to be more active than Sun-like stars.

\defcitealias{2014ApJ...783L..25M}{M14}

\subsubsection{HIP 67620}\label{67620_results}

This is a well-known binary and the target with the largest amount of RV data available (see Fig. \ref{orbits_updated}). Its orbital parameters have been previously determined by \citet{2006ApJS..162..207A} and more recently by \citet{2015MNRAS.453.1439J} and \citet{2016AJ....152...46W}. The orbital parameters we obtained are in good agreement with \citet{2016AJ....152...46W}. It has one of the most peculiar rotation rates from our sample (2.77 km s$^{-1}$ for an age of 7.18 Gyr), an enhanced chromospheric activity \citep{F16sub} and an anomalous [Y/Mg] abundance \citepalias{2016A&A...590A..32T}. The orbital period of the system is far too long for gravitational interaction to enhance the rotation of the main star through tidal acceleration, thus we should expect that they evolve similarly to single stars from this point of view.

High-resolution imaging of HIP 67620 revealed a companion with $V_{\mathrm{mag}} \approx 10$ and separations which are consistent with the spectroscopic companion (\citealt{2012AJ....143...42H}; \citetalias{2014AJ....147...86T}). As explained by \citet{2015MNRAS.453.1439J}, the presence of a companion with $m > 0.55$ M$_\odot$ can produce contaminations to the spectra that introduce noise to the measured RVs; our estimate of $m \sin{i}$ for this system is 0.58 M$_\odot$. These results suggest that, similarly to HIP 19911 but to a lesser degree, the companion of HIP 67620 may be offsetting our estimates for rotational velocity, stellar activity, chemical abundances and isochronal age.

We were unable to discern double-lines in the HARPS spectra, likely resulting from  unfavorable Doppler separations (observations range from February 2012 to March 2013). However, an analysis of the CCF of this star shows slight asymmetries in the line profiles of the HARPS spectra, which indicates a possible contamination by the companion. \citet{2017ApJ...836..139F} reported HIP 67620 as a double-lined binary using spectra taken at high-resolution ($R \approx 60$$,$$000$) in February 2014 and July 2015. As expected due to the short time coverage of the HARPS spectra, we did not see any correlation between the bisector inverse slope \citep[as defined in][]{2001A&A...379..279Q} and the radial velocities of HIP 67620.

\citet{2015MNRAS.453.1439J} found an additional signal on the periodogram of HIP 67620 at 532 d, which could be fit with a 1 M${_{\mathrm{Jup}}}$ planet, bringing down the $rms$ of the fit by a factor of 2. However, we did not find any significant peak in the periodogram of the residuals of the radial velocities for HIP 67620.

\subsubsection{HIP 103983}\label{103983_results}

The revised age for HIP 103983 ($4.9 \pm 0.9$ Gyr; Spina et al., in preparation) renders this system as an abnormally fast rotator ($3.38$ km s$^{-1}$) for its age. However, upon a careful inspection of the HARPS data obtained at different dates, we identified that the spectrum from 2015 July 27 displays clearly visible double-lines, albeit not as well separated as those observed in the HIRES spectra of HIP 19911 (see Fig. \ref{SBII}). No other anomalies besides enhanced rotation were inferred for this system. The CCF plots of the HARPS spectra show clear longer tails towards the blue side for most observations.

In \citetalias{2016A&A...592A.156D} we reported distortions in the combined spectra of HIP 103983; this is likely a result from the combination of the spectra at orbital phases in which the Doppler separation between the binaries is large. Since the observing windows of the HARPS spectra of HIP 19911 and HIP 67620 do not cover large RV separations (see Fig. \ref{orbits_updated}), the same effect is not seen in the combined spectra of these stars. This effect also explains why HIP 103983 is an outlier in fig. 4 of \citetalias{2016A&A...592A.156D}.

Although we have limited RV data, the \texttt{emcee} simulations converge towards a well-defined solution instead of allowing longer periods, as these produce larger residuals. It is important, however, to keep monitoring the radial velocities of this system in order to confirm that the most recent data points are in fact a second inflection in the radial velocities. The residuals for the fit for the HIRES spectra are on the order of 100 m s$^{-1}$, which is likely a result from the contamination by a bright companion. \citetalias{2014AJ....147...86T} reported a $0.91$ M$_\odot$ visual companion at a separation of $0.093 \arcsec$, which is consistent with the spectrocopic semi-major axis we estimated: $0.149 \arcsec$ for a distance of 65.7 pc \citep{2007A&A...474..653V}.

\subsection{Other binaries with updated orbital parameters}

\begin{figure*}
\centering
\begin{tabular}{cc}
\includegraphics[width=0.45\textwidth]{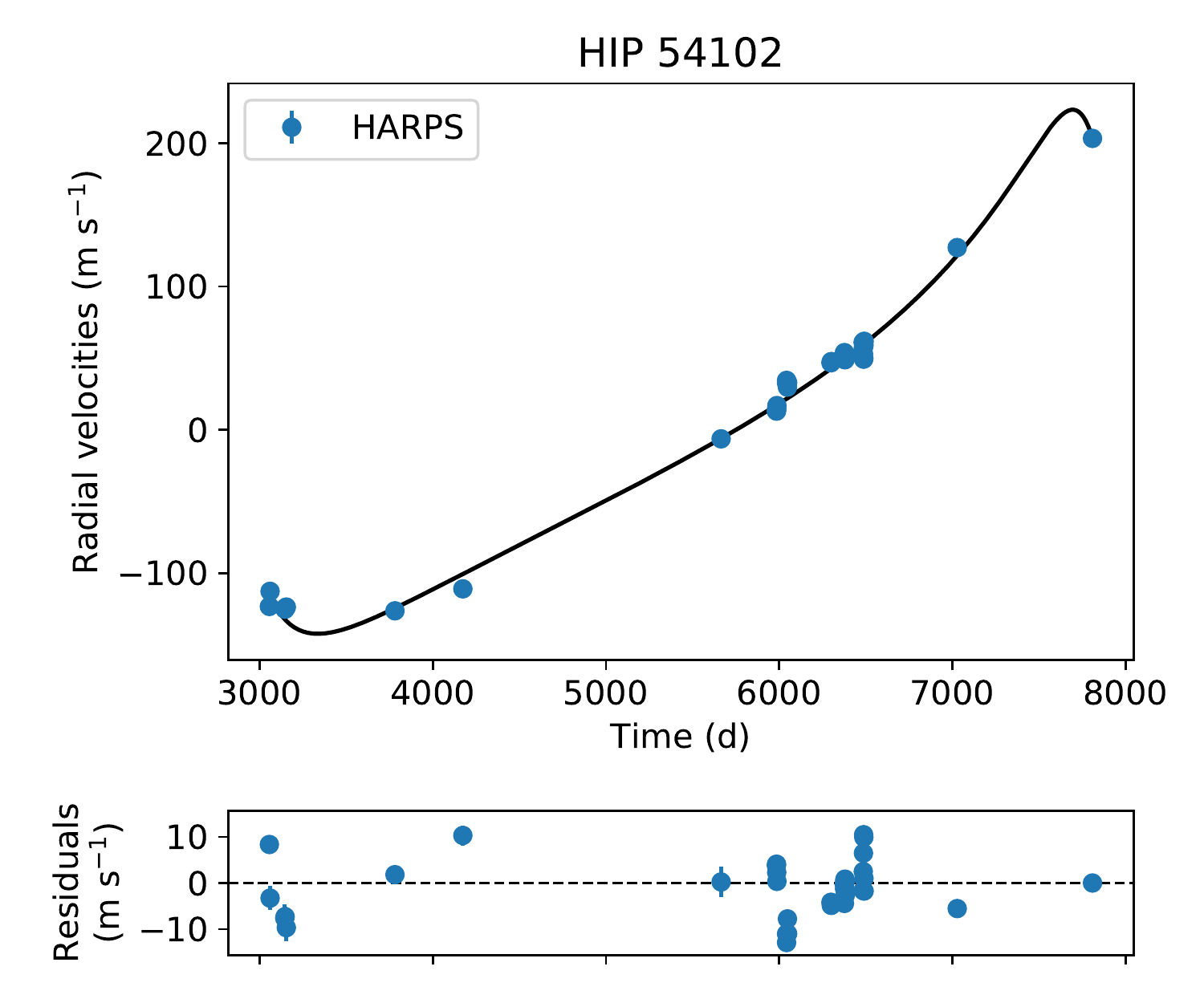} & \includegraphics[width=0.45\textwidth]{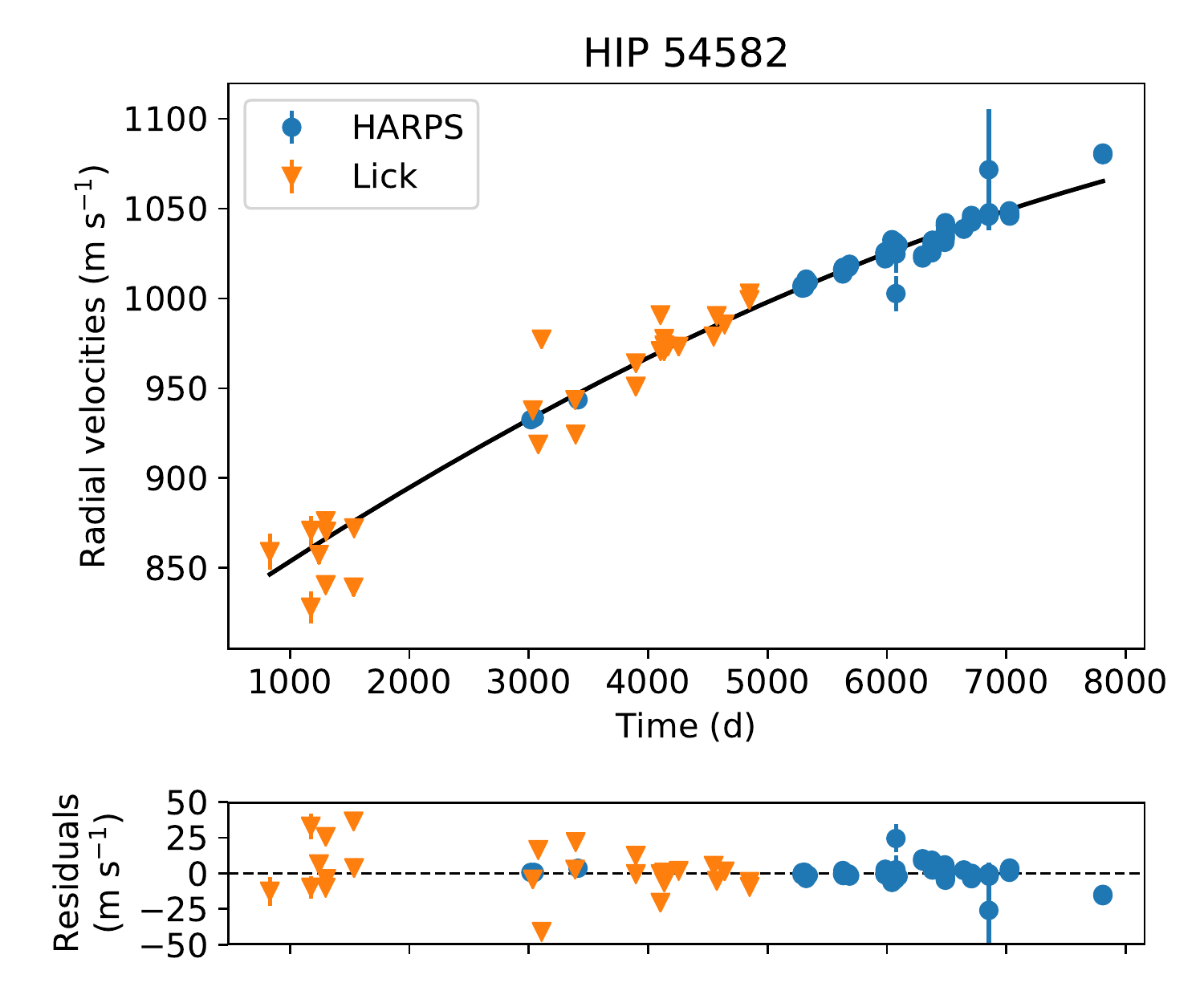} \\
\includegraphics[width=0.45\textwidth]{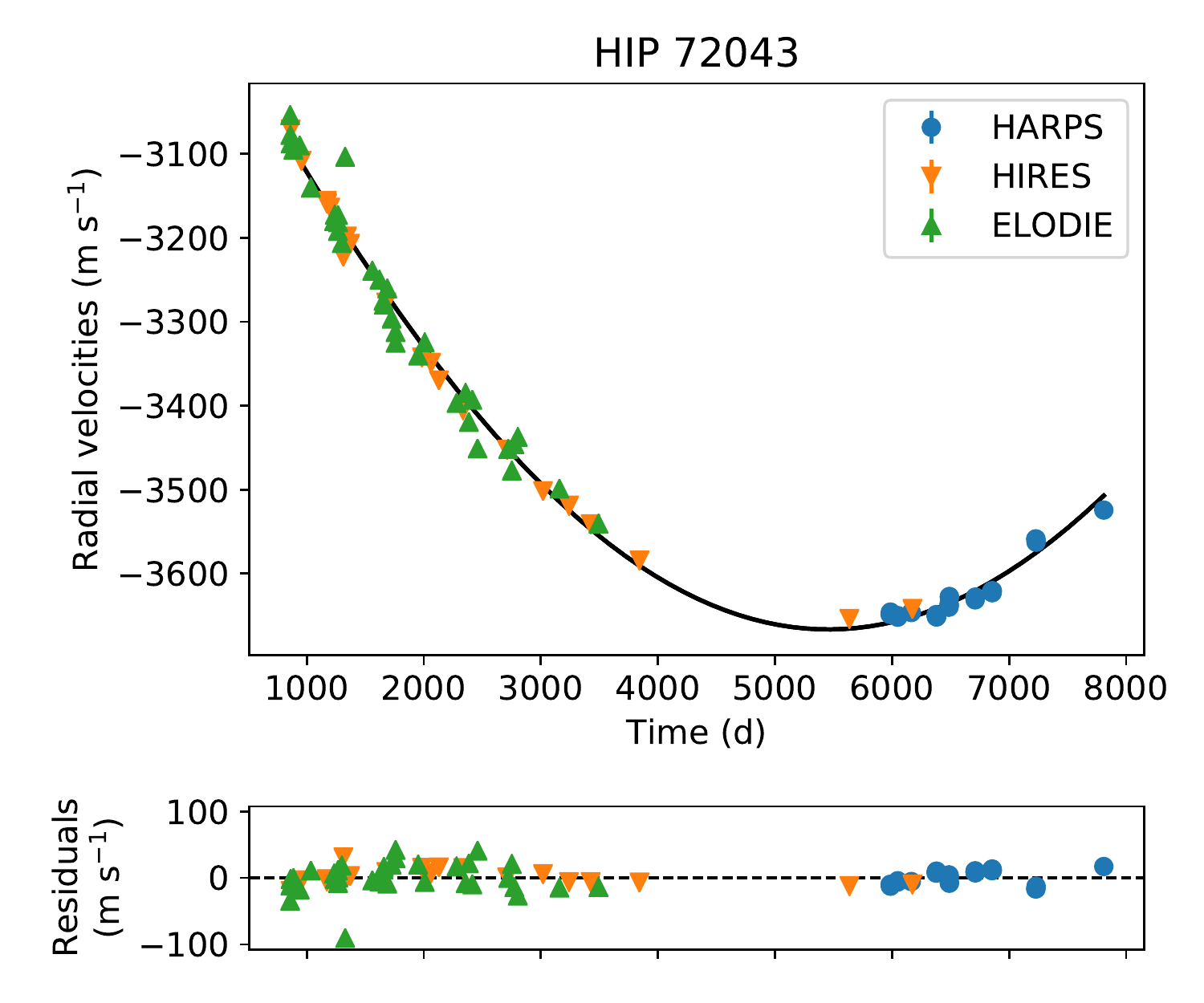} & \includegraphics[width=0.45\textwidth]{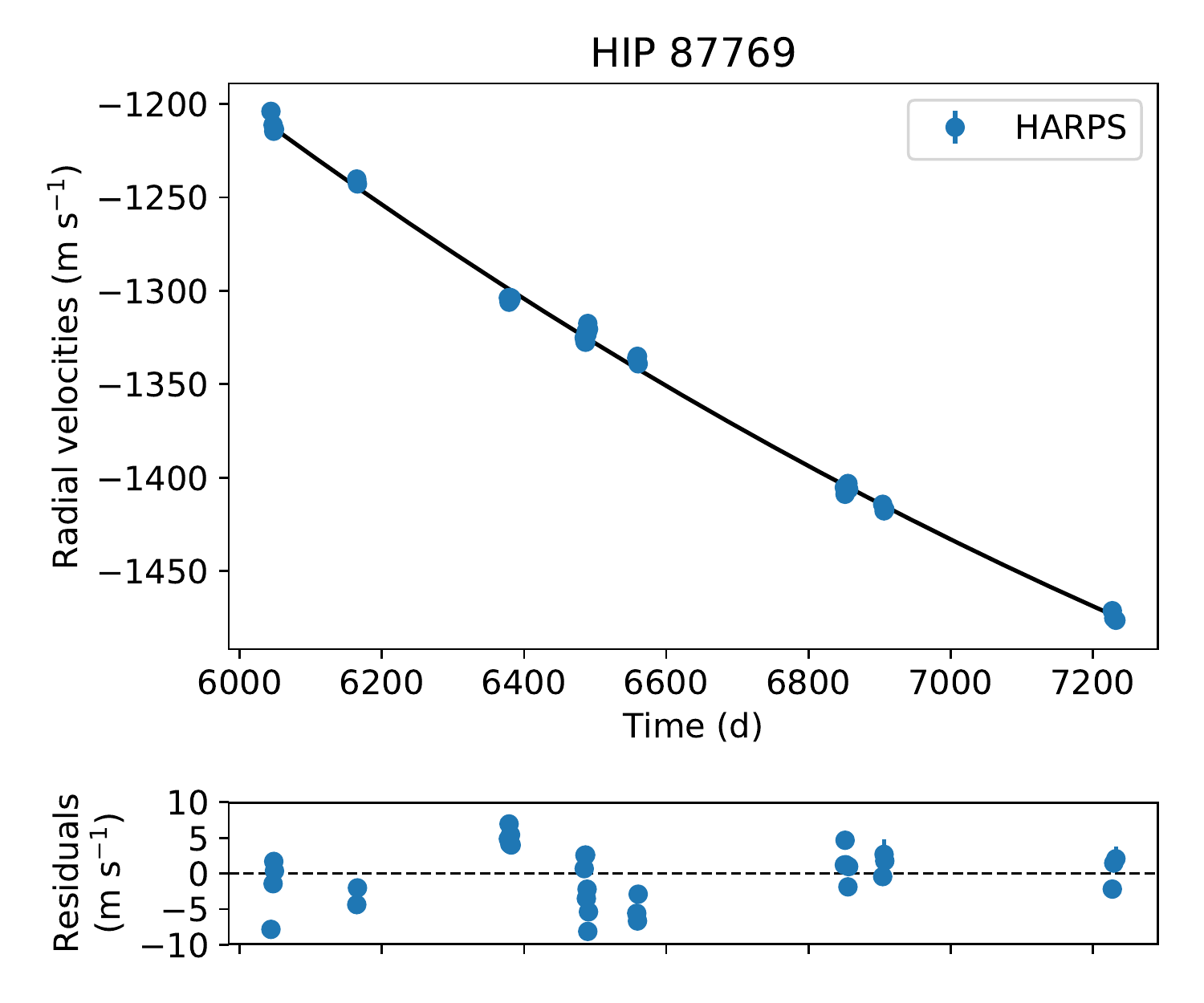} \\
\end{tabular}
\caption{The radial velocities and the solutions that produced the lower limits of orbital parameters of the solar twin binaries with RV curvature. A similar plot for HIP 73241 can be found in \citet{2015MNRAS.453.1439J}. Time is given in $\mathrm{JD} - 2.45 \times 10^{6}$ d.}
\label{long_period_rvs}
\end{figure*}

Among the known binaries in the solar twins sample, five of them display curvature in their RV data which allows the estimation of limits for their orbital parameters (see Table \ref{curvature_results} and Fig. \ref{long_period_rvs}). Some of the linear trend binaries observed in our HARPS Solar Twin Planet Search program are targets with large potential for follow-up studies. For the companions with visual detection, we were able to estimate their most likely mass (see Table \ref{linear_trend_results}).

\begin{figure}
\centering
\includegraphics[width=0.47\textwidth]{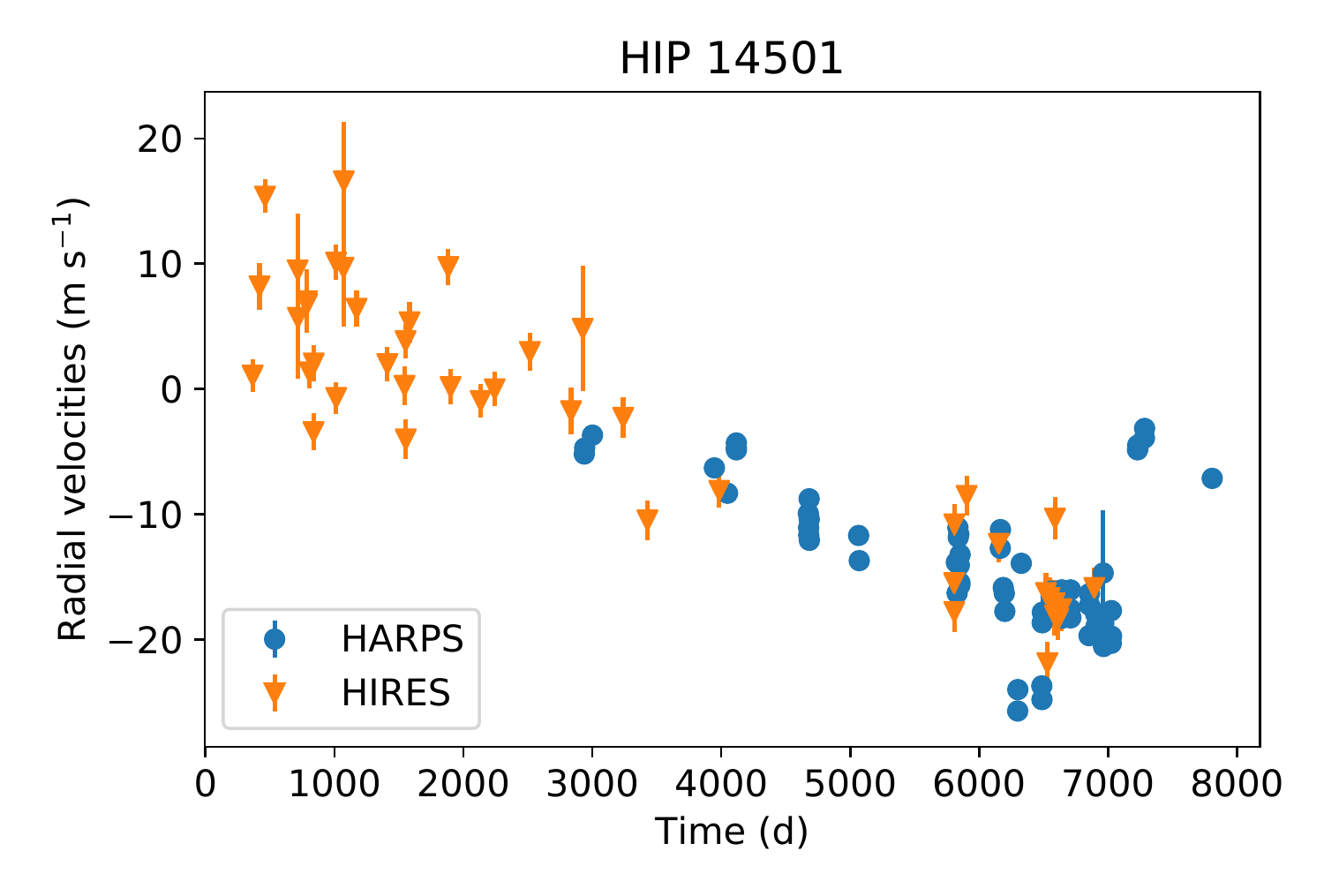}
\caption{Radial velocities of HIP 14501. The RV shift in the $y$-axes is arbitrary. Time is given in $\mathrm{JD} - 2.45 \times 10^{6}$ d.}
\label{rv_14501}
\end{figure}

\textbf{HIP 14501:} \textit{Linear trend.} Its companion is reported by \citet{2014ApJ...781...29C} as the first directly imaged T dwarf that produces a measurable doppler acceleration in the primary star. Using a low-resolution direct spectrum of the companion, \citet{2015ApJ...798L..43C} estimated a model-dependent mass of 56.7 M$_{\mathrm{Jup}}$. Using the HARPS and HIRES/Keck RV data and the observed separation of $1.653\arcsec$ \citep{2014ApJ...781...29C}, we found that the most likely value of the companion mass is 0.043 M$_\odot$ (45 M$_{\mathrm{Jup}}$), which agrees with the mass obtained by \citet{2015ApJ...798L..43C}. The most recent HARPS data hints of an inflection point in the orbit of HIP 14501 B (see Fig. \ref{rv_14501}), but further RV monitoring of the system is necessary to confirm it.

\textbf{HIP 18844:} \textit{Linear trend.} It is listed as a multiple system containing a closer-in low-mass stellar companion (estimated 0.06 M$\odot$, which agrees with our most likely mass) and orbital period $T = 6.5$ yr \citepalias[][and references therein]{2014AJ....147...86T}. For the companion farther away, \citet{2015MNRAS.453.1439J} reported a minimum orbital period of $\sim 195$ yr and $m\sin{i} = 0.33$ M$_\odot$, with a separation of $29\arcsec$ in 1941 ($\sim$$750$ AU for a distance of 26 pc).

\textbf{HIP 54102:} \textit{RV curvature only.} It is listed as a proper motion binary by \citet{2005AJ....129.2420M}, but there are no other information about the companions in the literature. Its eccentricity is completely unconstrained due to lack of RV coverage. We estimate that its companion's minimum mass is $12.6$ M$_\mathrm{Jup}$, with an orbital period larger than 14 years.

\begin{figure}
\centering
\includegraphics[width=0.47\textwidth]{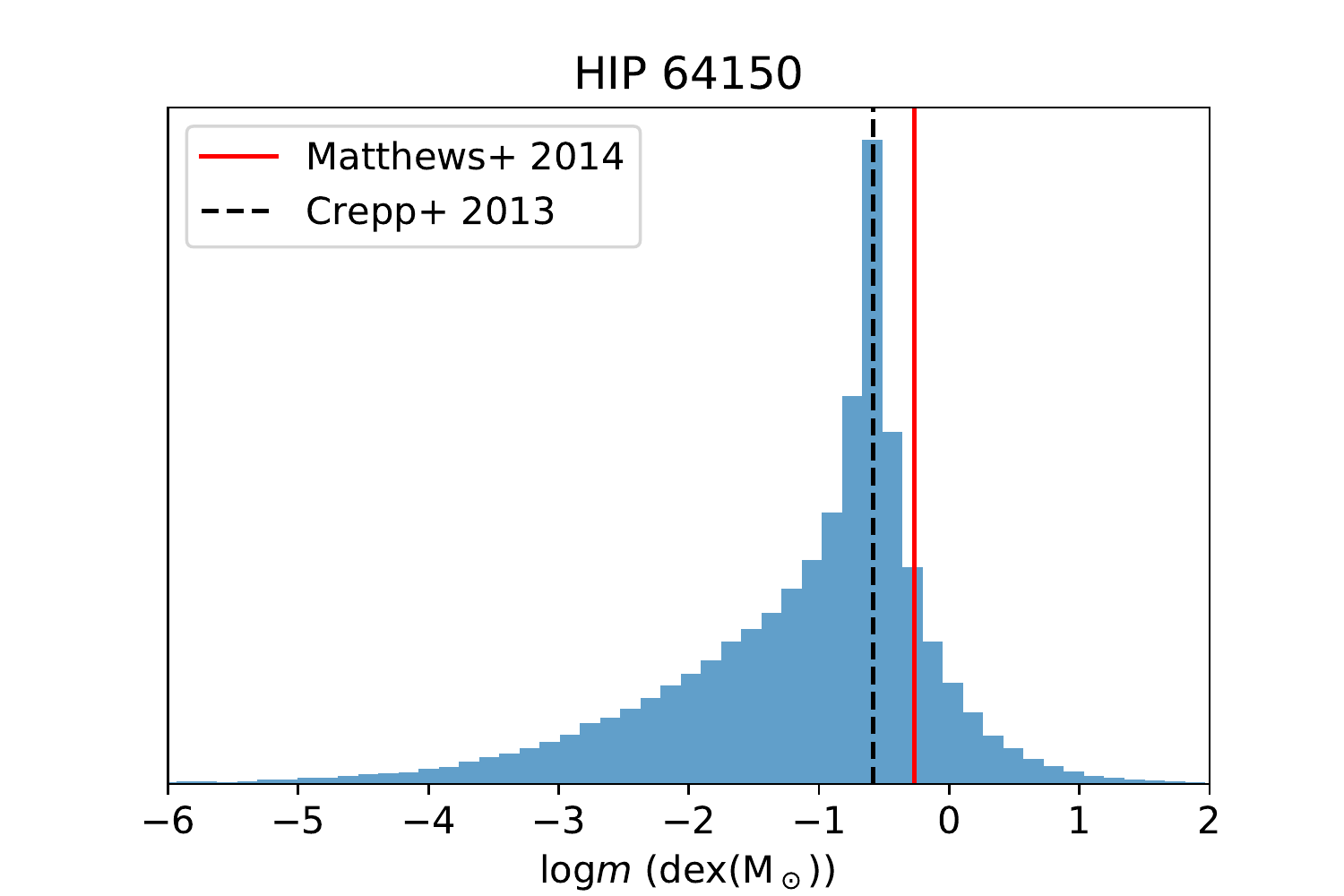}
\caption{Posterior probability distribution of the companion mass for HIP 64150. The mass obtained by \citet{2014ApJ...783L..25M} using SED fitting for the spectra of the WD companion is shown as a red vertical line.}
\label{64150_pdf}
\end{figure}

\textbf{HIP 64150:} \textit{Linear trend.} The most likely companion mass obtained by the method explained in Section \ref{methods_long_period} renders an estimate of 0.26 M$_\odot$, as seen in Fig. \ref{64150_pdf}. The higher mass (0.54 M$_\odot$) obtained by \citealt{2013ApJ...774....1C} and \citetalias{2014ApJ...783L..25M} can be attributed to less likely orbital configurations, but it is still inside the 1-$\sigma$ confidence interval of the RV+imaging mass estimate. The main star displays clear signals of atmosphere pollution caused by mass transfer from its companion during the red giant phase \citep{2011PASJ...63..697T}, characterizing the only confirmed blue straggler of our sample. The measured projected separation of the binary system is 18.1 AU \citepalias{2014ApJ...783L..25M}, which indicates that even for such a wide system the amount of mass transferred is still large enough to produce measurable differences in chemical abundances. It seems, however, that the amount of angular momentum transfer was not enough to produce significant enhancement in the rotation rate and activity of the solar twin. It is also important to note that the isochronal age measured for this system \citepalias{2016A&A...590A..32T} has a better agreement with the white dwarf (WD) cooling age estimated by \citetalias{2014ApJ...783L..25M} than previous estimates, illustrating the importance of studying these Sirius-like systems to test the various methods of age estimation.

\textbf{HIP 65708:} This star has previously been reported as a single-lined spectroscopic binary with an orbital solution \citep{2002AJ....124.1144L}. Here we update this solution by leveraging the extremely precise radial velocities measured in the Lick Planet Search program and with the HARPS spectrograph. The minimum mass of the companion is 0.167 M$_\odot$, indicating it is a red dwarf, orbiting at less than 1 AU with a slight eccentricity of 0.31. Our results agree with the previous orbital solution, which was based solely on data with uncertainties two orders of magnitude higher than the most recent data from HARPS and the Lick Planet Search.

\textbf{HIP 72043:} \textit{RV curvature only.} Similarly to HIP 54102, it is listed as a proper motion binary and we could not constrain its eccentricity. A fairly massive ($> 0.5$ M$_\odot$) companion is inferred at a very large period; this fit suggests that the longitude of periapse of the companion of HIP 72043 is currently at an unfavorable position for visual detection.

\textbf{HIP 73241:} \textit{RV curvature only.} The companion's orbit is eccentric enough to allow an estimation of the minimum eccentricity; its companion has been previously been confirmed by \citet{2010ApJS..190....1R} and visually detected by \citetalias{2014AJ....147...86T} with a separation $0.318\arcsec$. In \citetalias{2016A&A...592A.156D} we listed this star as having an unusually high rotation, but here we revise this conclusion and list HIP 73241 as a candidate peculiar rotator because its $v \sin{i}$ is less than 2$\sigma$ above the expected value for its age. Similarly to HIP 67620, this peculiarity, if real,  could also be explained by contamination by a bright companion, since we determined that the minimum companion mass $m \sin{i} > 0.49$ M$_\odot$.

\textbf{HIP 79578:} The companion is a well-defined 0.10 M$_\odot$ red dwarf orbiting the main star approximately every 18 years in a fairly eccentric orbit ($e = 0.33$). The orbital parameters we obtained differ significantly from the ones obtained by \citet{2015MNRAS.453.1439J} by more than $10\%$, except for the eccentricity; also in contrast, \citeauthor{2015MNRAS.453.1439J} report it as a brown dwarf companion. The fit for this binary displays residuals of up to 30 m s$^{-1}$ for the AATPS radial velocities, and the periodogram of these residuals shows a peak near the period 725 days. When we fit an extra object with $m\sin{i} = 0.70$ M$_{\mathrm{Jup}}$ at this period ($a = 1.62$ AU and $e = 0.87$), it improves the general fit of the RVs by a factor of 7. It is important to mention, however, that there are only 17 data points for the AATPS dataset, and the HARPS dataset does not display large residuals for a single companion fit. We need thus more observations to securely infer the configuration of this binary system, and if it truly has an extra substellar companion at a shorter period.

\textbf{HIP 81746:} This is another high-eccentricity ($e = 0.7$) binary that does not display clear anomalies in its rotation and activity. Its companion is a 0.1 M$_\odot$ red dwarf orbiting the main star every 9 years. The orbital parameters we obtained are in good agreement with the ones reported by \citet{2015MNRAS.453.1439J}.

\textbf{HIP 83276:} \textit{RV curvature only.} Although the HARPS radial velocities suggest the presence of a stellar mass companion, we do not have enough RV data points to infer any information about the orbital parameters of the system. Using radial velocities measured with the CORAVEL spectrograph, \citet{1991A&A...248..485D} found the companion has $m\sin{i}=0.24$ M$_\odot$, $e=0.185$ and an orbital period of 386.72 days.

\textbf{HIP 87769:} \textit{RV curvature only.} It is reported as a binary system by \citetalias{2014AJ....147...86T} but, similarly to HIP 54102, lacks an inflection point in its RV data from HARPS, which spans 3.3 yr. There is a wide range of possible orbital solutions that suggest $m \sin{i}$ varying from brown dwarf masses to $\sim 1$ M$_\odot$. Higher eccentricities ($e > 0.8$) can be ruled out as unlikely because they suggest a companion with $m \sin{i} \approx 1$ M$_\odot$ at an orbital period of more than 500 yr and $a > 80$ AU.

\begin{table}
\begin{center}
\caption{Lower limits of the orbital parameters of the spectroscopic binaries with curvature in their RV data.}
\begin{tabular}{llcccc}
\toprule[\heavyrulewidth]
\multirow{2}{*}{HIP} & \multirow{2}{*}{HD} & $K$ & $T$ & $m \sin{i}$ & $e$\\
 & & (km s$^{-1}$) & (yr) & (M$_\odot$) & \\
\bottomrule[\heavyrulewidth]
 54102 & 96116 & $> 0.182$ & $> 14$ & $> 0.012$ & $\dots$ \\
 54582 & 97037 & $> 0.193$ & $> 102$ & $> 0.03$ & $\dots$ \\
 72043 & 129814 & $> 2.11$ & $> 104$ & $> 0.40$ & $\dots$ \\
 73241 & 131923 & $> 5.93$ & $> 21.0$ & $> 0.49$ & $> 0.72$ \\
 87769 & 163441 & $> 1.90$ & $> 81.5$ & $> 0.30$ & $\dots$ \\
\bottomrule[\heavyrulewidth]
\label{curvature_results}
\end{tabular}
\end{center}
\end{table}

\begin{table}
\begin{center}
\caption{Measured RV slopes of the linear trend binaries. The most likely mass for the spectroscopic companion is estimated when their separation is available. Otherwise a minimum mass is provided.}
\begin{tabular}{llcccc}
\toprule[\heavyrulewidth]
\multirow{2}{*}{HIP} & \multirow{2}{*}{HD} & $dv_r / dt$ & $\rho$ & Dist. & $m$  \\
 & & (m s$^{-1}$ yr$^{-1}$) & (arcsec) & (pc)\textsuperscript{d} & (M$_{\mathrm{Jup}}$) \\
\bottomrule[\heavyrulewidth]
14501 & 19467 & $-1.30 \pm 0.01$ & 1.653\textsuperscript{a} & 30.86 & 45 \\
18844\textsuperscript{$\dag$} & 25874 & $424 \pm 3$ & 0.140\textsuperscript{b} & 25.91 & 79 \\
62039 & 110537 & $7.25 \pm 0.03$ & $\dots$ & 42.68 & $> 19$ \\
64150 & 114174 & $61.72 \pm 0.02$ & 0.675\textsuperscript{c} & 26.14 & 270 \\
\bottomrule[\heavyrulewidth]
\multicolumn{6}{l}{\textsuperscript{a}\footnotesize{\citet{2014ApJ...781...29C}.}}\\
\multicolumn{6}{l}{\textsuperscript{b}\footnotesize{\citet{2014AJ....147...86T}.}}\\
\multicolumn{6}{l}{\textsuperscript{c}\footnotesize{\citet{2014ApJ...783L..25M}.}}\\
\multicolumn{6}{l}{\textsuperscript{d}\footnotesize{\citet{2007A&A...474..653V}.}}
\\
\multicolumn{6}{l}{\textsuperscript{$\dag$}\footnotesize{Triple or higher-order system.}}\\
\label{linear_trend_results}
\end{tabular}
\end{center}
\end{table}

\subsection{Considerations on multiplicity statistics}

Although planet search surveys are generally biased against the presence of binaries due to avoiding known compact multiple systems, the fraction of binary or higher-order systems in the whole sample of the Solar Twin Planet Search program is $42\% \pm 6\%$\footnote{\footnotesize{Counting stellar and brown dwarf companions. The uncertainty is computed using a bootstrap resampling analysis with 10,000 iterations, similarly to \citet{2010ApJS..190....1R}. In each iteration, a new set of 81 solar twins is randomly drawn from the original sample allowing stars to be selected more than once.}}. This value agrees with previous multiplicity fractions reported by, e.g., \citet{2010ApJS..190....1R} and \citetalias{2014AJ....147...86T}; however, it is signficantly lower than the $58\%$ multiplicity factor for solar-type stars reported by \citet{2017ApJ...836..139F}, who argues that previous results are subject to selection effects and are thus biased against the presence of multiple systems.

\begin{figure}
\centering
\includegraphics[width=0.47\textwidth]{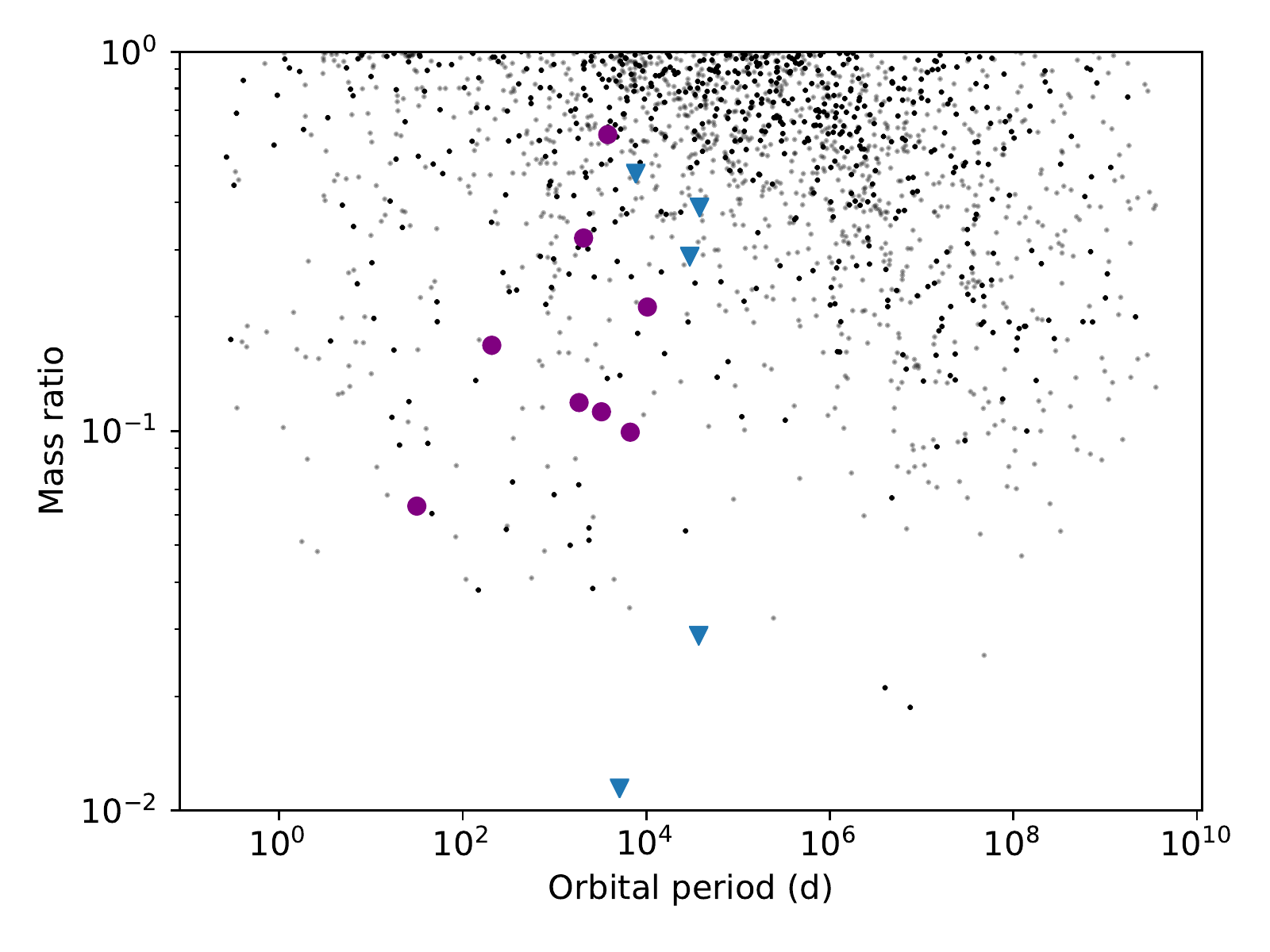}
\caption{Mass ratios in function of the orbital periods of binary stars or higher-order systems in the solar neighborhood. The purple circles are binaries in our sample with well defined period and $m \sin{i}$; the blue triangles correspond to the binaries in our sample for which we only have lower limits for the periods and $m \sin{i}$. The stars from \citetalias{2014AJ....147...86T} are plotted as black dots (the darker ones are those with main star masses between 0.9 and 1.1 M$_\odot$).}
\label{mratio}
\end{figure}

The orbital period vs. mass ratio plot of companions in the Solar Twin Planet Search is shown in Fig. \ref{mratio}. A comparison with the sample of solar-type stars from \citetalias{2014AJ....147...86T} reveals two important biases in our sample: i) Mass ratios are mostly below 0.3 because of selection of targets that do not show large radial velocity variations in previous studies; ii) Orbital periods are mostly lower than 30 yr because longer values cannot be constrained from the recent RV surveys targeting solar-type stars with low-mass companions. In such cases, further monitoring of linear trend and RV curvature-only binaries may prove useful to understand the origins of the brown dwarf desert \citep{2006ApJ...640.1051G}. These targets are particularly appealing because the long periods mean that the separation from the main star is large enough to allow us to observe them directly using high-resolution imaging.

Previous studies on the period-eccentricity relation for binary stars found that systems with orbital periods below 10 days tend to have eccentricities near zero, while those between 10 and 1000 days follow a roughly flat distribution of eccentricities \citep[][and references therein]{2016AJ....152..189K}, an effect that is due to the timescales for circularization of orbits. In relation to our sample, with the exception of HIP 30037, HIP 65708 and HIP 83276, all of the binaries we observed have periods longer than 1000 days and eccentricities higher than 0.3, which agrees with the aforementioned findings. According to \citet{1991A&A...248..485D}, the distribution of eccentricities on systems with $T > 1000$ d is a function of energy only, and does not depend on $T$ (see fig. 5 in \citeauthor{1991A&A...248..485D}). Interestingly, HIP 30037, which hosts a brown dwarf companion with $T = 31.6$ d, falls inside the 25--35 days interval of orbital periods found by \citeauthor{2016AJ....152..189K} that corresponds to a short stage of evolution of binaries undergoing a fast change in their orbits.

\section{Conclusions}

The Solar Twin Planet Search and several other programs observed 81 solar twins using the HARPS spectrograph. In total, 18 of these solar twins are spectroscopic binaries, 18 are visual binaries, and two intersect these categories. We found a multiplicity fraction of $42\% \pm 6\%$ in the whole sample, which is lower than the expected fraction ($\sim$$58\%$) because of selection effects that are generally seen in exoplanet search surveys.

We updated or reproduced the solutions of several known binaries, and determined all the orbital parameters of HIP 19911, HIP 65708, HIP 67620, HIP 79578, HIP 81746 and HIP 103983. The stars HIP 43297 and HIP 64673, which we previously reported as binaries, are likely to host long-period giant planets instead of stellar companions. For binaries with partial orbits, we were able to place lower limits for some of their orbital parameters owing to the presence of curvature or an inflection point in their RV data. We estimated the most likely mass of the companions of the binaries that display only linear trends in their RV data. Future work is needed on studying the long-period binaries using photometry data and high-resolution imaging in order to constrain the nature of their companions. These wide solar twin binaries are prime targets for detailed physical characterization of their companions owing to the favorable separation for AO imaging and the precision with which we can measure the stellar parameters of the main star -- this is particularly important for fully convective red dwarf stars and very low-mass companions such as the T dwarf HIP 14501 B, whose evolution and structure is still poorly constrained.

Additionally, we reported the detailed discovery of new companions to the following solar twins: HIP 6407, HIP 30037, HIP 54582, and HIP 62039, for which we are able to determine an orbital solution for the first two using radial velocities. The latter two do not have enough RV data to obtain precise orbital parameters, but we can nonetheless estimate their minimum companions masses. We found that these new companions are likely very low-mass, ranging from 0.02 to 0.12 M$_\odot$ (although stressing that these are lower limits), which should be useful in understanding the origins of the brown dwarf desert in future research.

The anomalies and RV residuals observed on HIP 19911, HIP 67620 and HIP 103983 are likely due to contamination by the companion on the spectra of the main star. Although the peculiar stars in our sample are no longer considered blue straggler candidates, it is important to note that the detection of WD companions is particularly important for the study of field Sun-like stars because they allow the estimation of their cooling ages; these are more reliable than isochronal and chromospheric ages in some cases, providing thus robust tests for other age estimate methods. We do not expect that the presence of M dwarf companions contaminate lithium spectral lines in Sun-like stars, thus stellar ages derived from Li abundances may be more reliable for double-lined solar twins. We recommend a revision of the stellar parameters of the peculiar binary stars by analyzing high-resolution spectra at the highest Doppler separations possible, or using Gaussian processes to disentangle the contaminated spectra \citep[see, e.g.,][]{2017ApJ...840...49C}.

We conclude that single-lined solar twin binaries with orbital periods larger than several months and moderate to low eccentricities do not display signals of distinct rotational evolution when compared to single solar twins. The most compact system in our sample, HIP 30037, which hosts a 0.06 M$_\odot$ brown dwarf companion at an orbital period of 31 days is, in fact, one of the quietest stars in the sample (in regards of its activity levels), and is thus a viable target for further efforts in detecting moderate- to long-period circumbinary planets.

\section*{Acknowledgements}

LdS acknowledges the financial support from FAPESP grants no. 2016/01684-9 and 2014/26908-1. JM thanks FAPESP (2012/24392-2) for support. LS acknowledges support by FAPESP (2014/15706-9). This research made use of SciPy \citep{scipy_ref}, Astropy \citep{2013A&A...558A..33A}, Matplotlib \citep{Hunter:2007}, and the SIMBAD and VizieR databases \citep{2000A&AS..143....9W, 2000A&AS..143...23O}, operated at CDS, Strasbourg, France. We thank R. P. Butler, S. Vogt, G. Laughlin and J. Burt for allowing us to analyze the LCES HIRES/Keck data prior to publication. LdS also thanks B. Montet, J. St\"urmer and A. Seifahrt for the fruitful discussions on the results and code implementation. We would also like to thank the anonymous referee for providing valuable suggestions to improve this manuscript.



\bibliographystyle{mnras}
\bibliography{bib}







\bsp	
\label{lastpage}
\end{document}